%% file: main.tex
\documentclass[journal]{IEEEtran}
\IEEEoverridecommandlockouts
\usepackage{cite}
\usepackage{amsmath,amssymb,amsfonts,amsthm,mathtools}

\usepackage{nccmath}
\usepackage[utf8]{inputenc}
\usepackage{mathtools,bbm} 
\usepackage{enumerate}
\usepackage{cases}
\usepackage{graphicx,epstopdf}
\usepackage{booktabs,multirow}
\usepackage{algorithmic}
\usepackage{tabularx}
\usepackage[ruled,vlined]{algorithm2e}
\usepackage[nolist]{acronym}
\usepackage{cuted}
\usepackage{color, soul}
\usepackage{subcaption}
\usepackage{scalerel}
\usepackage{orcidlink}
\usepackage{soul}
\setstcolor{red}

\newcolumntype{Y}{>{\centering\arraybackslash}X}
\newcolumntype{?}{!{\vrule width 1.5pt}}

\newcommand{\G}{\mathcal{G}}
\newcommand{\U}{\mathcal{U}}
\newcommand{\B}{B}
\newcommand{\xtg}{x_{t}^{(g)}}
\newcommand{\ytg}{y_{t}^{(g)}}
\newcommand{\xtu}{x_{t}^{(u)}}
\newcommand{\ytu}{y_{t}^{(u)}}

\newcommand{\ptg}{p_{t}^{(g)}}
\newcommand{\served}{\psi_{t}^{(g)}}

\newcommand{\T}{\mathcal{T}}

\input{acros}

\begin{document}
\bstctlcite{IEEEexample:BSTcontrol}

\title{Multi-Agent Meta-Advisor for UAV Fleet Trajectory Design in Vehicular Networks}
%

\author {Leonardo~Spampinato, Lorenzo~Mario~Amorosa, Enrico~Testi, Chiara~Buratti, Riccardo~Marini
\thanks{
Copyright (c) 2025 IEEE. Personal use of this material is permitted. However, permission to use this material for any other purposes must be obtained from the IEEE by sending a request to pubs-permissions@ieee.org
\textit{(Corresponding author: Leonardo~Spampinato.)}

Leonardo~Spampinato, Lorenzo~Mario~Amorosa, Enrico~Testi, Chiara~Buratti, and Riccardo Marini are with the National Laboratory of Wireless Communications of CNIT (Wilab, CNIT), Italy. 
(e-mail: leonardo.spampinato@unibo.it; lorenzomario.amorosa@unibo.it; enrico.testi@unibo.it; c.buratti@unibo.it; riccardo.marini@wilab.cnit.it).

Leonardo~Spampinato, Lorenzo~Mario~Amorosa, Enrico~Testi, and Chiara~Buratti are also with the Department of Electrical, Energy, and Information Engineering, University of Bologna, Italy.
}}

\maketitle
\begin{abstract}
Future vehicular networks require continuous connectivity to serve highly mobile users in urban environments. To mitigate the coverage limitations of fixed terrestrial macro base stations (MBS) under non line-of-sight (NLoS) conditions, fleets of unmanned aerial base stations (UABSs) can be deployed as aerial base stations, dynamically repositioning to track vehicular users and traffic hotspots in coordination with the terrestrial network.
This paper addresses cooperative multi-agent trajectory design under different service areas and takeoff configurations, where rapid and safe adaptation across scenarios is essential. We formulate the problem as a multi-task decentralized partially observable Markov decision process and solve it using centralized training and decentralized execution with double dueling deep Q-network (3DQN), enabling online training for real-world deployments.
However, efficient exploration remains a bottleneck, with conventional strategies like $\epsilon$-greedy requiring careful tuning. To overcome this, we propose the multi-agent meta-advisor with advisor override (MAMO). This framework guides agent exploration through a meta-policy learned jointly across tasks. It uses a dynamic override mechanism that allows agents to reject misaligned guidance when the advisor fails to generalize to a specific scenario.
Simulation results across three realistic urban scenarios and multiple takeoff configurations show that MAMO achieves faster convergence and higher returns than tuned $\epsilon$-greedy baselines, outperforming both an advisor-only ablation and a single generalized policy. Finally, we demonstrate that the learned UABS fleet significantly improves network performance compared to deployments without aerial support.
\end{abstract}


\begin{IEEEkeywords}
Meta learning, Vehicular networks, UAVs, Multi agent systems, Trajectory design
\end{IEEEkeywords}

\input{Chapters/1_Intro}
\input{Chapters/2_SOTA}
\input{Chapters/3_System}
\input{Chapters/DRL_System}
\input{Chapters/Exploration_Policy}
\input{Chapters/Numerical_Results}
\input{Chapters/8_Conclusions}
\appendix
\input{Chapters/4_RRM_system}

\section*{Acknowledgment}

This work has been carried out in the framework of the CNIT National Laboratory WiLab and the WiLab-Huawei Joint Innovation Center. We would like to thank Aman Jassal and Chan Zhou for the very fruitful discussion on this paper.

\bibliographystyle{IEEEtran}
\bibliography{ref}




\end{document}

%% file: acros.tex
\begin{acronym}

\acro{2D}{2nd dimension}
\acro{3D}{3rd dimension}
\acro{3DN}{3D networks}
\acro{3DQN}{double dueling deep Q-network}
\acro{3GPP}{3rd generation partnership project}
\acro{4G}{fourth-generation}
\acro{5G}{fifth-generation}
\acro{6G}{sixth-generation}
\acro{AOC}{advisor override count}
\acro{AI}{artificial intelligence}
\acro{AoI}{age of information}
\acro{APF}{artificial potential field}
\acro{ATG}{air-to-ground}
\acro{BS}{base station}
\acrodef{C2}{Command\&Control}
\acro{C2C}{command and control}
\acro{CA}{cooperative awareness}
\acro{CAM}{cooperative awareness message}
\acro{CAPEX}{capital expenditure}
\acro{CE}{coverage enhancement}
\acro{CoMPS}{continual meta policy search}
\acro{CP}{collective perception}
\acro{CPM}{collective perception message}
\acro{CPS}{collective perception service}
\acro{CSI}{channel state information}
\acro{CTDE}{centralized training decentralized execution}
\acro{D2D}{device-to-device}
\acro{DDPG}{deep deterministic policy gradient}
\acro{DDQN}{double deep Q-network}
\acro{Dec-POMDP}{decentralized partially observable Markov decision process}
\acro{DL}{downlink}
\acro{DNN}{deep neural network}
\acro{DPOMDP}{decentralized partially observable \ac{MDP}}
\acro{DQN}{deep Q-network}
\acro{DRL}{deep reinforcement learning}
\acro{ESN}{echo state network}
\acro{FDD}{frequency division duplexing}
\acro{FDMA}{frequency division multiple access}
\acro{FFS}{finite fourier series}
\acro{FoV}{field of view}
\acro{GoB}{grid of beams}
\acro{GNN}{graph neural network}
\acro{GU}{ground user}
\acro{GUE}{ground user equipment}
\acro{HAPS}{high altitude platform station}
\acro{IIoT}{industrial internet of things}
\acro{ILP}{integer linear program}
\acro{IoT}{internet of things}
\acro{ISAC}{integrated sensing and communication}
\acro{ISD}{inter-site distance}
\acro{ITS-S}{intelligent transport system station}
\acro{KPI}{key performance indicator}
\acro{LoS}{line-of-sight}
\acro{LSP}{large-scale parameter}
\acro{LTE}{long term evolution}
\acro{MAC}{medium access control}
\acro{MADRL}{multi-agent deep reinforcement learning}
\acro{MARL}{multi-agent reinforcement learning}
\acro{MAB}{multi-armed bandit}
\acro{MAML}{model agnostic meta learning}
\acrodef{MAMO}{multi-agent meta-advisor with advisor override}
\acro{MBS}{macro base station}
\acro{MDP}{markov decision process}
\acro{MEC}{mobile edge computing}
\acro{ML}{machine learning}
\acro{MLP}{multilayer perceptron}
\acro{mMTC}{massive machine-type communication}
\acro{MNO}{mobile network operator}
\acro{MSE}{mean squared error}
\acro{NB-IoT}{narrowband-internet of things}
\acro{NFZ}{no fly zone}
\acro{NLoS}{non line-of-sight}
\acro{NN}{neural network}
\acro{NOMA}{non-orthogonal multiple access}
\acro{O2I}{outdoor-to-indoor}
\acro{O2O}{outdoor-to-outdoor}
\acro{OFDM}{orthogonal frequency-division multiplexing}
\acro{OFDMA}{orthogonal frequency division multiple access}
\acro{OHE}{one hot encoding}
\acro{OMA}{orthogonal multiple access}
\acro{OPEX}{operational expenditure}
\acro{pmf}{probability mass function}
\acro{PPP}{poisson point process}
\acro{PRB}{physical resource block}
\acro{PSO}{particle swarm optimization}
\acro{QoE}{quality of experience}
\acro{QoS}{quality of service}
\acro{RB}{resource block}
\acro{ReLU}{rectified linear unit}
\acro{RF}{radio frequency}
\acro{RL}{reinforcement learning}
\acro{RRA}{radio resource assignment}
\acro{RRM}{radio resource management}
\acro{RSS}{received signal strength}
\acro{RSSI}{received signal strength indicator}
\acro{RSRP}{reference signals received power}
\acro{RSU}{road side unit}
\acro{RU}{resource unit}
\acro{SAR}{search and rescue}
\acro{SIC}{successive interference cancellation}
\acro{SINR}{signal to interference plus noise ratio}
\acro{SL}{supervised learning}
\acro{SNR}{signal-to-noise ratio}
\acro{SSB}{signal synchronization block}
\acro{SSP}{small-scale parameter}
\acro{SUMO}{simulation of urban mobility}
\acro{TBS}{terrestrial base station}
\acro{TCP}{thomas cluster process}
\acro{TD}{temporal difference}
\acro{TR}{technical report}
\acro{TSP}{traveling salesman problem}
\acro{UABS}{unmanned aerial base station}
\acro{UAS}{unmanned aerial system}
\acro{UAV}{unmanned aerial vehicle}
\acro{UE}{user equipment}
\acro{UL}{uplink}
\acro{UMa}{urban macro}
\acro{UMi}{urban micro}
\acro{UTM}{\ac{UAS} traffic management}
\acro{UUE}{unmanned aerial \ac{UE}}
\acro{V2C}{vehicle-to-cellular}
\acro{V2I}{vehicle-to-infrastructure}
\acro{V2R}{vehicle-to-roadside}
\acro{V2V}{vehicle-to-vehicle}
\acro{V2X}{vehicle-to-everything}
\acro{VO}{velocity obstacles}
\acro{FSE}{first successful episode}
\acro{mmWave}{millimeter wave}

\end{acronym}

%% file: Chapters/1_Intro.tex
\section{Introduction}
The evolution of 6G systems envisions support for \ac{V2X} communications, which demand stringent network requirements for safety-critical applications~\cite{Automated_Driving,Extended_Sensing_1,Extended_Sensing_2,11091498}.
However, the intrinsic high mobility of \acp{GUE}, coupled with spatio-temporal fluctuations in traffic density, presents unique obstacles to efficient network service~\cite{9195500,10.1145/1614269.1614278, 7470939}. 
In such dynamic scenarios, existing fixed \ac{MBS} deployments often fail to provide continuous connectivity. Signal paths are frequently obstructed by buildings, resulting in severe \ac{NLoS} conditions and coverage gaps that degrade service quality~\cite{GIORDANI2018158,9954321, 10595097, 11230595}. 
To mitigate these issues, the use of \acp{UABS} has emerged as a promising solution. 
By acting as aerial integrated access and backhaul relays, \acp{UABS} can provide on-demand capacity boost~\cite{8999435,8713514,10666852}.
The effectiveness of a \ac{UABS} fleet is determined by its dynamic positioning strategy. Leveraging their unmatched mobility, they can continuously adjust their trajectories to track vehicular users and anticipate service demand hotspots, both in time and space, while coordinating with the underlying terrestrial \ac{MBS} for backhauling and interference management through joint terrestrial-aerial \ac{RRM}. 
Thus, the problem of fleet trajectory design becomes critical.
The use of the \ac{MADRL} algorithm is envisioned to enable the fleet to learn an adaptive trajectory planning model, allowing \acp{UABS} to autonomously cooperate to guarantee strict \ac{V2X} \ac{QoE} requirements.

However, in real-world deployment, the fleet may be required to serve distinct service areas, each with a given road layout and traffic density, or be dispatched from diverse takeoff positions within the same area.
A robust design solution must possess the capability to adapt its operational behavior across these diverse scenarios without requiring a computationally expensive redesign each time a new operational scenario is introduced. 
Achieving such rapid adaptation poses a significant challenge, particularly for state-of-the-art \ac{MADRL} approaches. 
These methods typically require extensive retraining whenever deployment conditions change, relying on inefficient random exploration strategies~\cite{8903067,LI2026113489,9387125}.
In the context of a \ac{UABS}-assisted \ac{V2X} network, the resulting transient performance degradation caused by this retraining is prohibitive. Consequently, developing mechanisms for rapid and safe adaptation across multiple deployment scenarios
is fundamental for real-world viability.
To address these challenges, and building upon our previous work in~\cite{11011207}, we propose the \ac{MAMO} framework. \ac{MAMO} introduces a meta-advisor that guides the exploration phase of multi-agent systems. 
Rather than relying on random exploration strategies, agents receive guidance from a meta-policy that is learned jointly across multiple fleet configurations, providing generalization across the considered set of tasks.
By aggregating experience across operation scenarios, the advisor processes agents’ local observations to suggest exploration actions that accelerate per-task adaptation.
The framework further incorporates a dynamic override mechanism that enables agents to evaluate the advisor’s recommendations against their own task-specific policies. By allowing agents to reject task-mismatched guidance, e.g., when the advisor becomes over-confident toward only a subset of tasks, \ac{MAMO} ensures rapid adaptation to new environments while remaining robust to unique local conditions that a generalized advisor may fail to capture. Overall, \ac{MAMO} provides a safer and more efficient learning paradigm, significantly reducing the time-to-proficiency in multi-task adaptation.
The main contributions of this paper are summarized as follows:
\begin{itemize}
    \item we define a multi-task formulation for cooperative \ac{UABS} fleet trajectory design within a \ac{V2X} network, accounting for integrated terrestrial-aerial \ac{RRM} and adaptation requirements across diverse operational environments;
    \item we introduce a \ac{MADRL} algorithm utilizing a \ac{CTDE} architecture that exploits available network links to enable online training capabilities for the fleet, ensuring the system is applicable to real-world deployment scenarios;
    \item we propose the \ac{MAMO} exploration framework to drive effective adaptation via enhanced exploration among diverse tasks, directly improving the time-to-proficiency and operational readiness of the trained agents;
    \item we validate our framework through extensive simulations in realistic urban scenarios, demonstrating significant gains in training efficiency and convergence speed across a wide range of task settings compared to standard approaches.
\end{itemize}

The remainder of the paper is organized as follows: 
in Section~\ref{sec:soa}, the state-of-the-art on multi-agent exploration strategies is reported; Section~\ref{sec:system_model} presents the considered system model; Section~\ref{sec:MADRL_system} introduces the fleet trajectory design problem formulation and the adopted \ac{MADRL} system;
Section~\ref{sec:exploration} describes the exploration strategies investigated in our work, and details the proposed \ac{MAMO} solution;
Section~\ref{sec:results} reports numerical results, and in Section~\ref{sec:conclusions}, conclusions are drawn. 

%% file: Chapters/2_SOTA.tex
\section{Related Work}
\label{sec:soa}
The efficacy of \ac{MADRL} systems training relies on the exploration strategy adopted, especially in high-dimensional and dynamic environments such as \ac{UABS} fleet trajectory planning~\cite{10021988,10994242,drones7040236}. Conventional approaches, such as $\epsilon$-greedy, can be inefficient or require careful tuning, which is impractical for online learning in deployed networks~\cite{dabney2021temporallyextended,NEURIPS2023_2a91de02, 9682129, 10391701}. 
Accordingly, recent \ac{MADRL} literature has investigated mechanisms to improve exploration efficiency and stability, including intrinsic motivation, meta-learning, role-awareness, and teacher/advisor paradigms.
In the following, we review these directions and discuss their limitations in \ac{UABS}-aided vehicular networks, where learning should be fast, safe, and robust to heterogeneous deployment conditions.

Curiosity-driven exploration is widely used to address sparse rewards by augmenting the learning signal with intrinsic exploration bonuses. 
In~\cite{kim2024strangeness}, a strangeness-driven mechanism is proposed that promotes exploration of novel joint states, while~\cite{wang2022multi} extends prediction error–based intrinsic rewards to local agent observations, improving exploration under partial observability. However, these approaches require task-specific reward tuning, limiting their applicability in heterogeneous deployments, and auxiliary reward shaping can degrade network performance during training.
In contrast, our approach avoids the need for auxiliary intrinsic reward by delegating exploration to a separate advisor that is meta-trained to generalize across various tasks and promoting cross-task adaptability.

Meta-reinforcement learning and advisor-based exploration aim to transfer knowledge across a task distribution, thus accelerating adaptation. In \cite{Garcia2019AMA}, an external advisor is learned in a meta-\ac{MDP}, while \cite{ganapathi2022multi} studies how agents can selectively follow advice based on its trustworthiness for the current task. A key practical limitation is that many advisor approaches rely on separate offline pre-training phases, which may reduce online applicability. In contrast, our advisor is a modular component decoupled from the exploitation policy. It can be trained in parallel with the agents and online, using its own algorithm and hyperparameters. In this work, we implement it as a value-based off-policy learner to exploit experience replay and improve sample-efficiency across tasks.

Other works improve exploration by learning agent roles or estimating agent influence. In \cite{hu2021roma}, ROMA discovers emergent roles that guide exploration, while \cite{mahajan2019maven} introduces MAVEN, which leverages latent variables to encourage policy diversity. Similarly, \cite{du2021liir} proposes an influence-based mechanism that promotes exploration by rewarding agents with higher team impact. While these approaches enhance decentralized exploration diversity, they often suffer from brittle role assignments or unstable learning under dynamic task distributions. In contrast, our advisor is trained explicitly to handle variability in the environment configuration, like service maps and drone takeoff positions, without relying on latent role discovery or manually shaped influence models. 

Some works enhance exploration by augmenting the learning signal through representation learning. In \cite{gupta2021contrastive}, contrastive learning is used to obtain diverse feature encodings, while \cite{zheng2022representation} proposes predictive models to learn informative state abstractions. While often beneficial, these methods typically increase training complexity, requiring auxiliary losses and additional modules, and may require a broad set of solvable tasks to generalize in a reliable way.
In comparison, our method maintains simplicity and modularity by leveraging an augmented state representation that includes context features, i.e., simple known task descriptors, thus eliminating the need for learnable auxiliary predictions or models.
In summary, most existing approaches rely on (i) modifying reward functions with intrinsic signals, (ii) enforcing structural biases via role definitions, or (iii) increasing complexity through representation learning. While effective in static environments, the computational overhead and rigidity of these methods often limit their utility in scenarios requiring efficient online training.
To overcome these limitations without adding structural complexity to the agents themselves, we introduce a multi-agent meta-advisor. This modular component learns a generalized exploration policy across diverse scenarios, enabling multi-task adaptation. Because the advisor is decoupled from the agent's core architecture, it can be trained online with independent hyperparameter tuning, making it ideal for \ac{UABS} fleet trajectory design, where fast and safe convergence is critical for service continuity.

%% file: Chapters/3_System.tex
\section{System Model}
\label{sec:system_model}
\subsection{Reference Scenario}
\begin{figure}[t]
    \centering
    \includegraphics[width=0.95\linewidth, trim=0 60 80 20, clip]{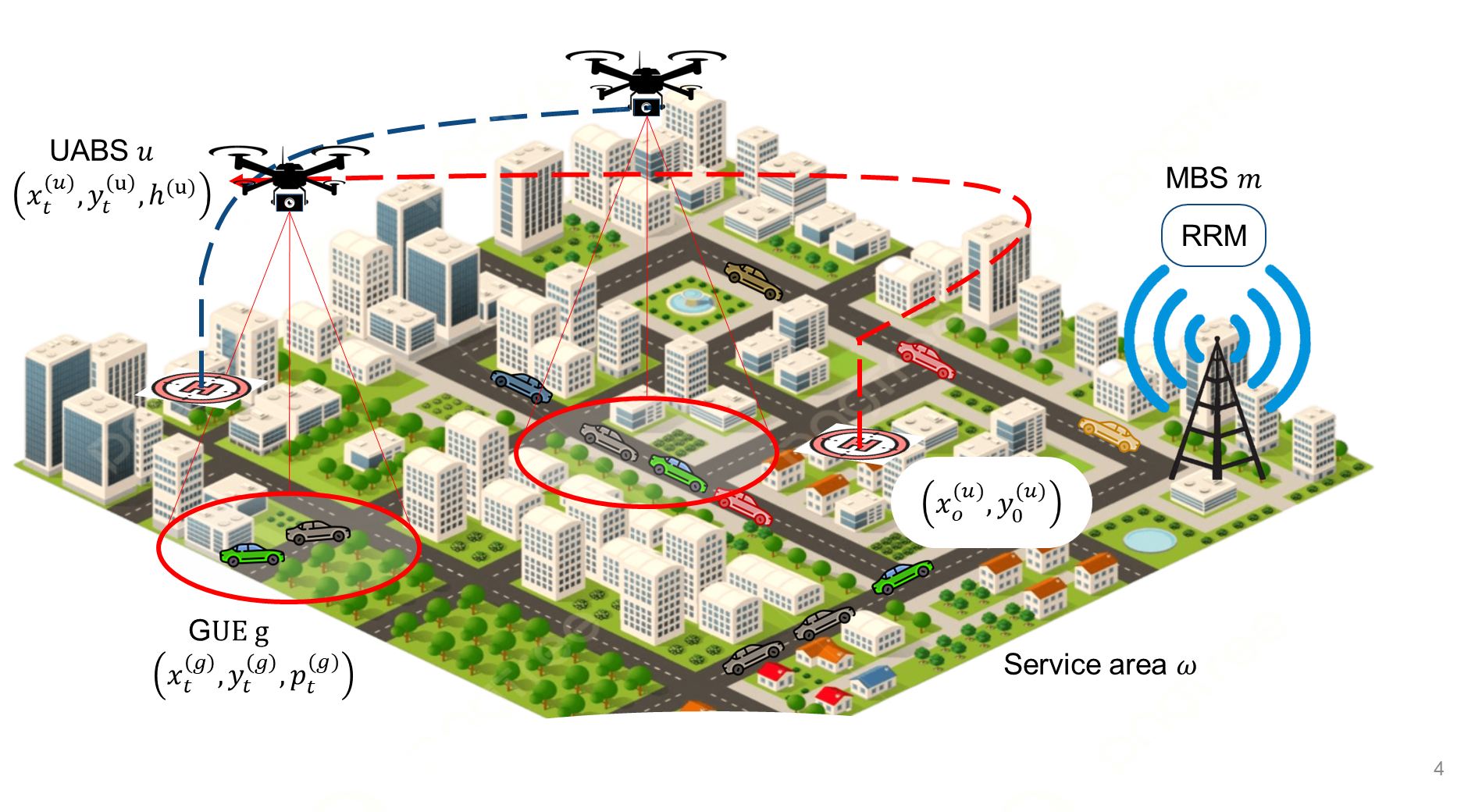}
    \caption{Representation of the reference scenario.}
    \label{fig:scenario}
\end{figure}
We consider an urban scenario in which a set of \acp{GUE} denoted by $\G$, participates in a \ac{V2X} application. Each \ac{GUE} $g\in\G$ periodically generates and uploads data packets to the network within a designated service area $\omega\in\Omega$, where $\Omega$ denotes the set of service areas considered. Each area is characterized by a specific street layout map, within which users travel according to a specific road traffic distribution. 
Cellular coverage within $\omega$ is provided by a terrestrial \ac{MBS}, denoted by $m$, operating at carrier frequency $f_c$ in \ac{mmWave}. The location of the \ac{MBS} is given by $(x_\omega^{(m)}, y_\omega^{(m)})$, where this notation highlights the \ac{MBS} dependence on the specific service area $\omega\in\Omega$. 

To complement the fixed terrestrial infrastructure and mitigate its limitations in terms of coverage and capacity, a fleet of \acp{UABS}, denoted by the set $\U$, is deployed in each area. The \acp{UABS} act as aerial base stations, providing additional access opportunities to the \acp{GUE}, aiming at enhancing the overall system performance. Each \ac{UABS} $u\in\U$ is assumed to move at a flying speed $v^{(u)}$ at a fixed altitude $h^{(u)}$. However, their trajectory is not given a priori.
Each \ac{UABS} performs a flight mission of duration $T$ which consists of a sequence of discrete timesteps $t = 0, 1, \dots, T$, each of duration $\Delta t$. At each time step, the positions of the \acp{UABS} are represented as $(\xtu, \ytu)$ $\forall u \in \mathcal{U}$.
The ordered sequence $\mathbf{P_t}=\big( \xtu, \ytu \;\big|\; u\in\mathcal{U}\big)$ represents the positions of all \acp{UABS} in the fleet at time step $t$, with $\mathbf{P}_0$ denoting their takeoff coordinates.

Each \ac{UABS} maintains a wireless backhaul link to the \ac{MBS}, ensuring continuous connectivity with the terrestrial infrastructure. Through this link, user traffic from \acp{GUE} served by \acp{UABS} is relayed to the core network, while operational commands are exchanged to coordinate \ac{UABS} behavior. This backhaul connection is essential for seamless integration of \acp{UABS} operations within the network.
Since \acp{UABS} operate within the same frequency band as the terrestrial network, their deployment requires no additional spectrum allocation for mobile operators. However, this shared spectrum approach demands careful interference management to maintain network performance.
To address this challenge, resource management is centralized at the \ac{MBS}, which acts as the controller for both terrestrial and aerial network segments. A \ac{RRM} algorithm running at the \ac{MBS} dynamically manages user association and radio resource allocation, determining whether each \ac{GUE} should connect to the \ac{MBS} or to a \ac{UABS}. 
The algorithm maximizes the continuity of service of served users, accounting for potential interference.

Within this scenario, our goal is to determine flight mission fleet trajectories $\mathbf{P}_{0:T,\omega}=[\mathbf{P}_0,\mathbf{P}_1,\dots,\mathbf{P}_T]_\omega$ 
that maximize the overall network performance in coordination with the \ac{MBS} through joint aerial-terrestrial \ac{RRM}, for each service area $\omega$ and takeoff locations $\mathbf{P}_0$ considered. 
A representation of the considered scenario is shown in Fig.~\ref{fig:scenario}.

\subsection{Reference Application}\label{subsec:ref.application}
Users move within the considered service area $\omega \in \Omega$ based on its street layout and traffic distribution, with their time-varying position denoted as $(\xtg, \ytg)$. Each \ac{GUE} $g\in\G$ periodically generates \acp{CAM} messages of fixed size $D_g$, requesting radio resources for the transmission. This setup corresponds to an extended sensing scenario, where \acp{GUE} continuously offload locally gathered sensor data, ranging from telemetry to high-definition video feeds, to the network infrastructure~\cite{ganesan20195g}. 

Within the service area, a \ac{GUE} $g$ is said to be \textit{covered} by a \ac{BS}, where a \ac{BS} is either the \ac{MBS} or a \ac{UABS}, if the corresponding wireless link among the two terminals achieves a \ac{SINR} above a predefined threshold $\text{SINR}_{\rm th}$. 
We model this using the binary variables $c_t^{(g,m)}$ and $c_t^{(g,u)}$, which are equal to 1 if $g$ is covered by the \ac{MBS} $m$ or by a \ac{UABS} $u\in\U$, respectively. These variables are determined based on the \ac{SINR} computation detailed in Sec.~\ref{subsec:channel_model}, and depend primarily on the positions between \acp{GUE}, \ac{MBS}, and \acp{UABS} in the fleet. 

Given the network coverage state, which allows to estimate the achievable rates, a centralized \ac{RRM} algorithm determines the optimal user association and uplink resource scheduling at each time step. To this end, a \ac{GUE} $g$ is considered \textit{served} at time step $t$ if the \ac{RRM} algorithm has assigned sufficient resources to upload its data packet, as part of its optimization output process. 
This is represented by the binary variables $\psi_t^{(g)}$, which equals 1 when $g$ is served and 0 otherwise. We further define $\psi_t^{(g,m)}$ and $\psi_t^{(g,u)}$ to indicate whether it has been served by \ac{MBS} $m$ or a \ac{UABS} $u\in\U$, respectively. 
A user can be served by at most one \ac{BS}. 
Details on the \ac{RRM} algorithm formulation are reported in Appendix~\ref{sec:RRM_model}.

To capture the continuous service requirements typical of vehicular applications, we define a \ac{GUE} $g$ as \textit{satisfied} if it is served for at least $\hat{N}_s$ time steps within its service window $W_g$. Each service window $W_g$ comprises $N_{\rm w}$ time steps and lasts for $T_{\rm w}=N_{\rm w}\Delta t$.
The \ac{QoE} requirement $\hat{N}_s$ reflects the time a vehicle may take to maneuver, such as turning at intersections or roundabouts. Example values can be found in~\cite{5GAA}.

To track the current service continuity status at each time step $t$, a priority metric $\ptg$ is assigned to each \ac{GUE} $g$ to count the number of times each $g$ has been served in its current service window $W_g$.
In particular, within each service time window $W_g$ (i.e., at time steps $t=1,..,N_{\rm w}$) $p_{t}^{(g)}$ varies as follows:
\begin{equation}\label{ch:single_agent,eq:pg}
p_{t}^{(g)} =
\begin{cases}
1, & \text{if } t=1,\\
p_{t-1}^{(g)}+1, & \text{if } \served=1,\\
p_{t-1}^{(g)},   & \text{if } \served=0.
\end{cases}
\end{equation}
This priority is used as the primary maximization objective for the \ac{RRM} algorithm, enforcing continuous service and satisfaction of the \ac{QoE} constraints, but it also affects the fleet trajectory learning through observations and reward definitions, as explained in Sec.~\ref{sec:MADRL_system}.

\subsection{Channel Model}
\label{subsec:channel_model}
The channel model follows the \ac{UMa} channel described in the 3GPP TR 38.901~\cite{TR38901}. 
The adopted channel model differentiates between \ac{LoS} and \ac{NLoS} propagation regimes. This distinction is based on the \ac{LoS} probability $\rho_L$, which is defined as a function of the distance between the \ac{GUE} $g$ and the \ac{BS} (\ac{UABS} or \ac{MBS}).
The path loss variation due to shadowing can be described through a log-normal distribution with zero mean and with a standard deviation $\sigma_{\rm{LoS}}$ and $\sigma_{\rm NLoS}$ for \ac{LoS} and \ac{NLoS}, respectively.
Consequently, the received power and the \ac{SINR} can be derived as: 
\begin{equation}
    P_{\rm rx} = \frac{P_{\rm tx}G_{\rm rx}G_{\rm tx}}{PL}, \quad \text{SINR} = \frac{P_{\rm rx}}{P_{\rm noise}+ \sum_{i=1}^{N_{\rm int}}P_{\text{rx}, i}}\,,
\end{equation}
where $P_{\rm tx}$ is the transmitted power, $G_{\rm tx}$ and $G_{\rm rx}$ are the transmitter and receiver antenna gains, 
$PL$ is the path loss calculated following \mbox{Table~7.4.1-1} and~\mbox{7.4.2-1} in~\cite{TR38901}, $P_{\rm noise}$ the noise power, $N_{\rm int}$ denotes the number of interferers and $P_{\text{rx}, i}$ is the received power from the $i$-th interferer. The achievable data rate depends on the \ac{SINR} through the Shannon capacity formula. 

For what concerns the modelling of the \ac{UABS} beams, by defining $\phi^{(u)}$ as the field of view of the \ac{UABS} on the vertical plane, and $\Phi^{(u)}= 2\pi(1-cos(\phi^{(u)}/2))$, the corresponding solid angle, then the solid angle of the single beam may be approximated as $\Phi^{(u)}_{\rm beam} \approx \Phi^{(u)}/N_{\rm beam}$. Finally, the receiving gain of \ac{UABS} can be expressed as follow~\cite{beamforming_book}: 
\begin{equation}
G_{\rm beam}=\frac{41000}{\big(\Phi^{(u)}_{\rm beam}\frac{360}{2\pi}\big)^2}.    
\label{eq;beamforming}
\end{equation}
For the sake of simplicity, it is assumed that the radiation pattern of the equivalent beam is ideal, with gain $G_{\rm beam}$ inside $\Phi^{(u)}_{\rm beam}$ and 0~dB outside, i.e. \acp{GUE} that are not inside a beam are not considered connected to the \ac{UABS}, thus this assumption provides a worst-case scenario for what concerns the drone and vehicles connectivity. 

%% file: Chapters/DRL_System.tex
\section{Problem Formulation}
\label{sec:MADRL_system}
The stochastic nature of the service environment, characterized by unknown user positions within a service area $\omega \in \Omega$, dynamic channel variability, and the impact of the \ac{RRM}, makes the user state evolution analytically complex. 
As a consequence, determining the optimal fleet-wide mission trajectory $\mathbf{P}_{0:T,\omega}$ with direct and closed-form optimization is extremely complicated.
This problem can be framed as a sequential decision-making problem under uncertainty, allowing the use of \ac{MADRL} algorithms, supported by a \ac{CTDE} framework, to solve it.
This approach enables each \ac{UABS} to act as an autonomous agent, executing a decentralized control policy that maps real-time observations to trajectory decisions. 
To enable efficient learning and coordination within the fleet, the policy is learned as one unique model, shared across the entire fleet, and updated centrally during training to ease the burden of computation.

To achieve this, agents must go through an extensive learning phase of trial-and-error exploration. However, the efficiency of this exploration process represents a critical bottleneck. Conventional exploration strategies often fail to navigate the vast joint state-action space, leading to prohibitively slow convergence. This limitation is particularly evident in environments characterized by heterogeneous scenarios, where the inability to generalize across diverse tasks makes naive exploration inefficient.
To address this, we reduce the number of episodes required to reach target performance by employing a meta-advisor, which is trained over the full set of operational scenarios and aims at improving exploration strategies. Indeed, through the use of meta-exploration, the fleet can learn effective policies in significantly less time, enabling online and safe training for \ac{UABS} networks in real-world applications.

\subsection{Decentralized Partially Observable MDP}
To address the fleet trajectory design problem, we model it as a \ac{DPOMDP} with elements $(\mathcal{S},\mathcal{O},\mathcal{A},s_0,\mathcal{P},\mathcal{R})$, which provides the mathematical framework for multi-agent sequential decision-making under uncertainty. 
We consider a set of \ac{UABS} trajectory design problems exhibiting structural similarities. They are referred to as \textit{tasks} $\tau_i$, all belonging to the task set $\mathcal{T}$. Each task is uniquely defined by its operational scenario, $\tau_i=(\omega_i, \mathbf{P}_{0,i})$, representing a specific service area $\omega_i$ and fleet initial takeoff positions $\mathbf{P}_{0,i}$. 
%
While all tasks share the same global state space $\mathcal{S}$, observation space $\mathcal{O}$, and action space $\mathcal{A}$, each task $\tau_i$ constitutes a distinct instance of the \ac{DPOMDP}. Specifically, every task is characterized by a unique initial state distribution $s_{0,\tau_i}$, transition function $\mathcal{P}_{\tau_i}$, and reward function $\mathcal{R}_{\tau_i}$.

The interaction loop for a given generic task $\tau_i$ is as follows. At each time step $t$, the environment is in a global state $s_t \in \mathcal{S}$, which is not directly accessible to the agents.
The initial state $s_0$ is determined by the task's takeoff positions $\mathbf{P}_0$. Each \ac{UABS} $u \in \mathcal{U}$ perceives a local and limited observation $o_t^{(u)} \in \mathcal{O}$. 
This observation is a vector $o_t^{(u)}=\left[ \xtu, \ytu, t, \mathbf{P}_t, \mathbf{b}_t^{(u)} \right]$, where: (i) $\xtu, \ytu$ the current \ac{UABS}'s position; (ii) $t$ the current mission time step; (iii) $\mathbf{P}_t$ the position vector of all the \acp{UABS} in the fleet; and (iv) $\mathbf{b}_t^{(u)}$ is the \textit{per-beam information} that is a vector of size $N_{\rm beam}$ whose elements $b_{i,t}$ corresponds to the sum of priorities $p_{t}^{(g)}$ for all \acp{GUE} $g$ served by the $i$-th beam of \ac{UABS} $u$ at time $t$. Without loss of generality, the global state $s_t$ can be regarded as the union of all agents' local observations. 

Based on this local observation, each agent $u$ selects a discrete movement action $a_t^{(u)}$ from the action space $\mathcal{A}=\left[\leftarrow, \uparrow, \rightarrow, \downarrow, \nwarrow, \nearrow, \searrow, \swarrow, \emptyset \right]$. This set includes eight directions of movement and the hovering in place action $\emptyset$. The combination of all individual actions defines the global action $a_t$.

The joint action $a_t$ applied to the global state $s_t$ produces a state transition according to the task-specific dynamic and transition probability function $\mathcal{P}_{\tau_i}$, defining the probability $\mathcal{P}_{\tau_i}(s,a)=P(s_{t+1}|s_t,a_t)$ to transition to a new state $s_{t+1}$.

As a result of this transition, all agents receive a shared and global reward $r_t=r_t^{(u)} ,\ \forall u\in\mathcal{U}$ based on the task-specific reward function $\mathcal{R}_{\tau_i}(s,a,s')$.
This reward is defined as:
\begin{equation}
    r_t = r_t^{(u)} = \frac{\sum_{u\in\mathcal{U}}\sum_{g\in\mathcal{G}}\psi_t^{(g,u)} p_t^{(g)}}{\vert \mathcal{U} \vert}
\end{equation}
which represents the average number of users served by the whole fleet weighted by their priority $p_t^{(g)}$ at time $t$. 
The reward function is computed based on the centralized \ac{RRM} module, which is treated as an integral part of the environment dynamics. For a given fleet configuration, the \ac{RRM} module determines association and scheduling (see Appendix~\ref{sec:RRM_model}), defining which \acp{GUE} are served and the resulting reward. This service-based metric captures the inherent aerial-terrestrial coordination. Moreover, it is task-specific because each task $\tau_i$ defines a unique service area $\omega_i$ with specific \acp{GUE} and \ac{MBS} distributions, so identical state-action pairs may yield different service outcomes $\psi_t^{(g)}$ depending on the operational context.



Finally, agents will perceive the new state $s_{t+1}$ via their next local observation $o_{t+1}$, completing the interaction loop at time $t$. 

The agents' behavior within a given task $\tau_i$ is governed by a decentralized, task-specific policy $\pi_i(a_t^{(u)}|o_t^{(u)})$, mapping local observations to a probability distribution over actions. 
For each task $\tau_i \in \mathcal{T}$, the objective is to learn an optimal policy $\pi_i^*$ that maximizes the expected episodic cumulative discounted reward, that is:
\begin{equation} 
\pi_i^* = \arg\max_{\pi_i} \mathbb{E}_{\pi_i, \mathcal{P}_{\tau_i}}
\left[ \sum_{t=0}^{T} \gamma^t r_t \right] 
\label{eq:optimal_policy}
\end{equation}
with $\gamma\in (0,1)$ the discount factor, balancing the importance of immediate and future rewards.
The expectation is taken over all possible state-action pairs generated by agents following their policy $\pi_i$ in the task environment $\tau_i$ with transition functions $\mathcal{P}_{\tau_i}$ and reward function $\mathcal{R}_{\tau_i}(s,a,s')$.
Since all agents receive a shared reward, optimizing this objective induces the collaborative fleet behavior. 
Thus, the optimal mission fleet trajectory $\mathbf{P}_{0:T,\omega_i}$ corresponds to the sequence of \acp{UABS} positions resulting from agents following policy $\pi^*_i$ at each time step $t$.

\subsection{Multi-Agent Deep Reinforcement Learning System}
To find the optimal policy $\pi_i^*$ and solve the problem defined by Eq.~\eqref{eq:optimal_policy} for each task $\tau_i$, we employ the \ac{3DQN} algorithm. \ac{3DQN} is a state-of-the-art, off-policy \ac{DRL} algorithm that improves upon the vanilla \ac{DQN}~\cite{DQN} by integrating the Double DQN principle~\cite{Double_DQN} and the Dueling Network architecture~\cite{dueling_dqn}.
The algorithm aims to learn the optimal action-value function, $Q_i^*(o, a)$, for each task $\tau_i$. This function, $Q_{\pi_i}$, for a given policy $\pi_i$, observation $o_t$, and action $a_t$ is defined as the expected discounted return:
\begin{equation}
Q_{\pi_i}(o,a)=\mathbb{E}_{\pi_i}\left[ \sum_{k=t}^{T-1} \gamma^{k-t} r_k \, \Big| \, o=o_t, a=a_t\right]
\end{equation}
This represents the expected cumulative reward achieved by taking action $a_t$ in observation $o_t$ and subsequently following policy $\pi_i$ until the end of the episode.

For each task $\tau_i$, \ac{3DQN} employs two neural networks: an \textit{online} network with parameters $\theta_i$ and a \textit{target} network with parameters $\theta_i^-$. These networks receive a per-agent observation $o_t^{(u)}$ as input and output the estimated Q-value for each possible action $a \in \mathcal{A}$. 
The optimal actions $\overline{a}_t^{(u)}$ are then selected according to:
\begin{equation}
\label{eq:optimal_action}
    \overline{a}_t^{(u)} = \arg\max\limits_{a} Q_{\pi_i}(o_t^{(u)}, a|\theta_i)
\end{equation}

Furthermore, networks implement a dueling architecture, which decomposes the Q-value into a state-value $V(o)$ and a state-action advantage $A(o,a)$. The advantage represents the relative advantage of each action given the current observation, leading to improved learning performance.

To update the network parameters, all agents $u$ operating on task $\tau_i$ collect experience tuples $(o_t^{(u)}, a_t^{(u)},r_t,o_{t+1}^{(u)})$ at each time step $t$, storing them in a task-specific replay buffer $\mathcal{K}_i$. 
At each training iteration, a batch of $|k|$ experiences is randomly drawn from $\mathcal{K}_i$.
We denote the $e$-th experience tuple in this batch as $(o_e,a_e,r_e,o'_e)$, where $o'_e$ corresponds to the $o_{t+1}^{(u)}$ from the original tuple.
The online parameters $\theta_i$ are updated via gradient descent:
\begin{equation}
\theta_i \gets \theta_i - \alpha\nabla_{\theta_i} L(\theta_i)
\end{equation}
where $\alpha$ is the learning rate and $L(\theta_i)$ is the mean-squared error loss, defined using the Double DQN target $y_e$:
\begin{equation}
\label{eq:loss}
    \begin{aligned}
    L(\theta_i) &= \frac{1}{|k|} \sum_{e=1}^{|k|} \left( y_e - Q(o_e, a_e \mid \theta_i) \right)^2 \\
    \text{where} \quad y_e &= r_e + \gamma Q(o'_e, a' \mid \theta_i^-) \\
    a' &= \arg\max_{a} Q(o'_e, a \mid \theta_i)
    \end{aligned}
\end{equation}
This target $y_e$ decouples the action selection, provided by the online network $\theta_i$, from the value estimation, which uses the target network $\theta_i^-$ for its estimation, stabilizing learning. Every $Y$ training iterations, the online network parameters are copied to the target network: $\theta_i^- \gets \theta_i$.

\begin{figure}
    \centering
    \includegraphics[width=0.9\linewidth, trim=10 10 10 10, clip]{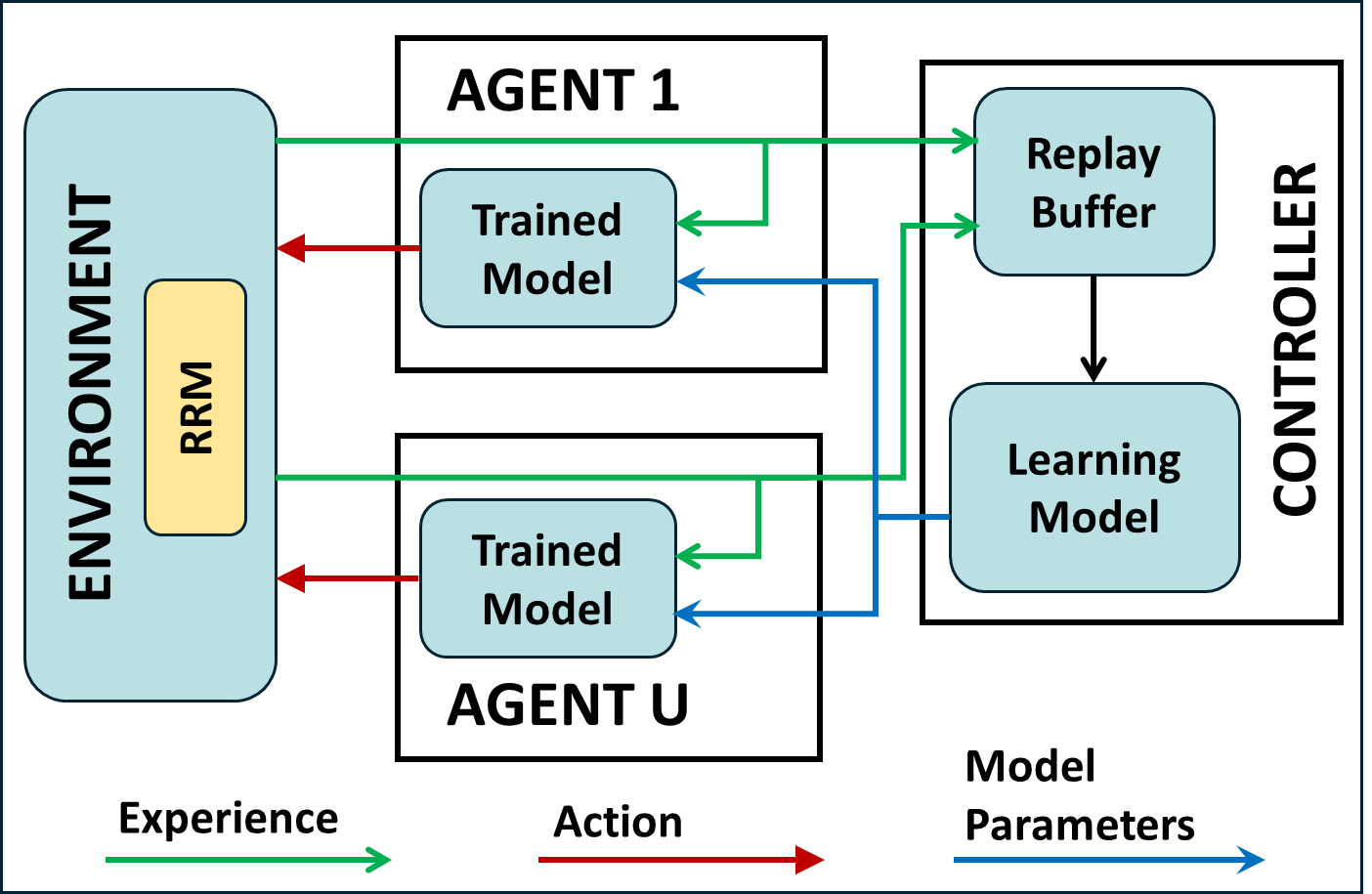}
    \caption{\ac{CTDE} architecture for UABS-aided vehicular networks.
    Each UABS $1, \ldots, U$ has a shared local copy of the trained model, which is updated periodically by the controller entity from experiences gathered by all the autonomous agents. 
    }
    \label{fig:CTDE}
\end{figure}

This \ac{MADRL} system is supported by the \ac{CTDE} architecture shown in Fig.~\ref{fig:CTDE}, 
which allows agents to train a model in a centralized manner, with agents performing inference of the learned model in a distributed way.
Each \ac{UABS} acts as a distributed agent, interacting continuously with the environment by performing actions based on local observations. Each experience tuple is then forwarded to the controller. The controller holds the shared replay buffer $\mathcal{K}_i$ for each task, and it is in charge of training the learning models $\theta_i$. The updated online parameters $\theta_i$ are then forwarded to each agent. This architecture enables system efficiency by reducing the computational burden on each agent.
It is assumed that the controller is co-located with the terrestrial \ac{MBS}, ensuring that the exchange of experiences and models is performed leveraging the backhaul link used for user traffic. 
For the sake of simplicity, this overhead communication is not accounted for in the \ac{RRM} procedure; however, its effect on the whole network performance was investigated in~\cite{11257456}.
Leveraging this architecture, where the learning loop is embedded within the network infrastructure, our framework establishes a practical basis for an online training procedure.
Here, online means that learning models are updated while the system is in execution, i.e., not relying on offline pre-training, leveraging the existing backhaul connectivity to exchange experiences and model updates.
In this paper, we validate the proposed online learning architecture through simulations; nevertheless, the same \ac{CTDE} loop can be executed over an operational network.
Online training is particularly relevant in complex urban deployments, where the actual road traffic distribution, channel conditions, and interference patterns may differ significantly from those assumed in offline simulators, causing model mismatch~\cite{9606868}.
However, deploying learning agents in a live network introduces a critical performance requirement: time-to-proficiency, i.e., how quickly agents converge to a useful policy without prolonged periods of poor service.
To address this, we propose the \ac{MAMO} system, which provides guided exploration and drastically reduces time-to-proficiency, making online multi-task adaptation quicker and safer.

%% file: Chapters/Exploration_Policy.tex
\section{Exploration Policies}
\label{sec:exploration}
\begin{figure*}[t!]
\centering
\begin{subfigure}[t]{1\columnwidth}
    \centering
    \includegraphics[width=0.9\columnwidth, trim= 50 0 0 0, clip]{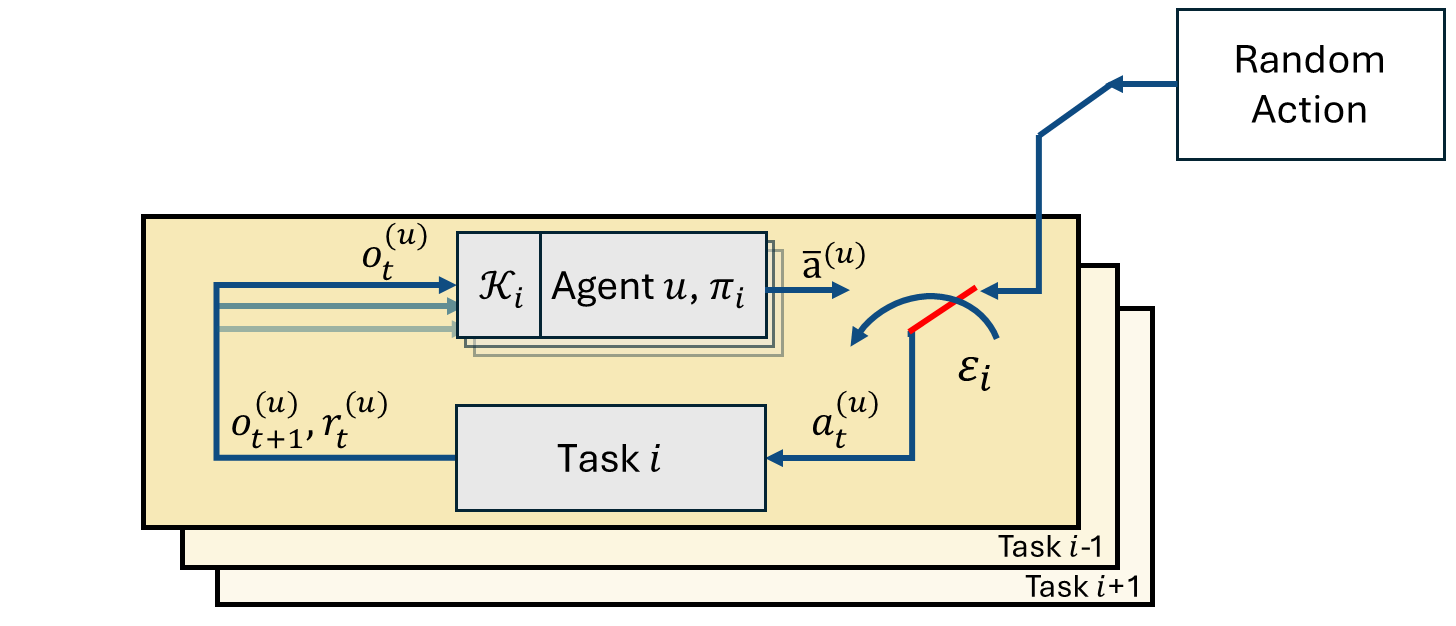}
    \caption{Individual training with random and independent exploration.}
    \label{fig:epsilon-greedy}
\end{subfigure}
\hfill
\begin{subfigure}[t]{1\columnwidth}
    \raggedleft
    \includegraphics[width=0.9\columnwidth]{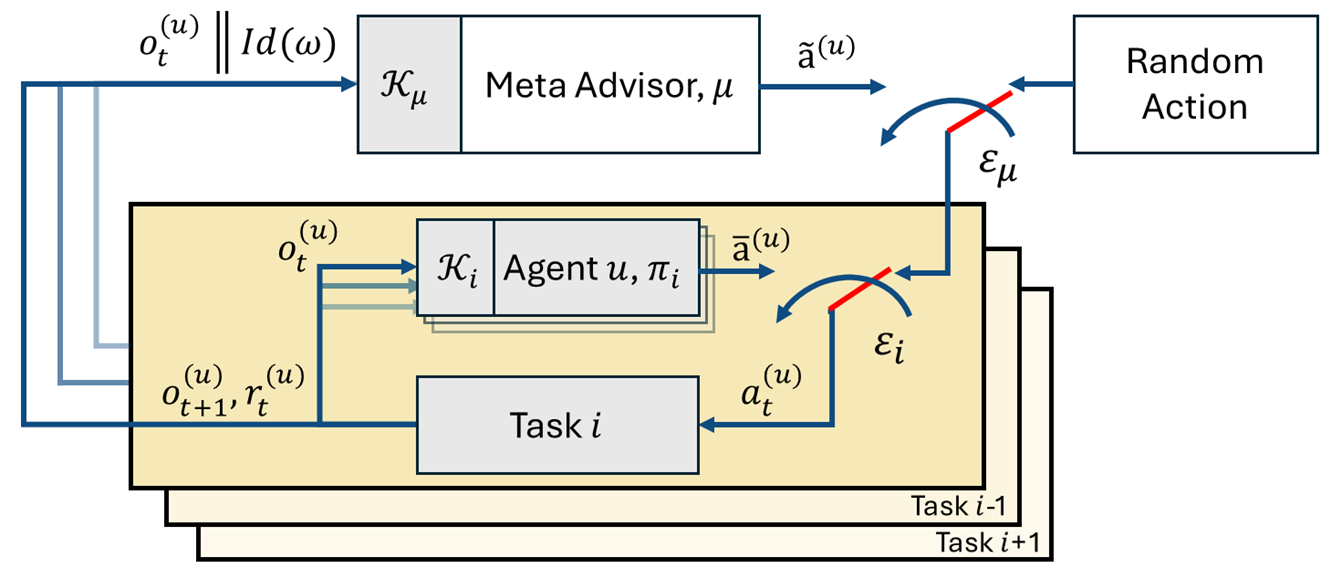}
    \caption{Proposed multi-agent meta-advisor architecture.}
    \label{fig:MAMA}
\end{subfigure}
\caption{Exploration strategies for multi task training.}
\end{figure*}

When dealing with \ac{MADRL} algorithms, each agent must balance the exploration-exploitation tradeoff to train efficiently. Early in training, 
when policy parameters are affected by random initialization,
exploration facilitates the collection of experiences $(o_t^{(u)},a_t^{(u)},r_t^{(u)},o_{t+1}^{(u)})$, guiding model updates.
However, inefficient or overly prolonged exploration can limit the exploitation of learned knowledge, leading to suboptimal convergence and longer training~\cite{NEURIPS2023_2a91de02}. In contrast, an efficient exploration allows agents to characterize the task-specific dynamics, quickly maximizing the expected cumulative reward. In this work, we compare two exploration strategies for the \acp{UABS}'s fleet, a classical and a meta-learning driven approach.
\subsection{\texorpdfstring{$\epsilon$-greedy}{epsilon-greedy}}
In this widely adopted baseline~\cite{sutton}, each agent $u\in\U$ selects a random action with probability $\varepsilon_i$ and with probability 1-$\varepsilon_i$ chooses the exploitation action $\overline{a}_t^{(u)}$ based on the current observation $o_t^{(u)}$, as defined in Eq.\eqref{eq:optimal_action}, according to the current policy parameters $\theta_i$.
Throughout a training of $N$ episodes, $\epsilon_i$ is linearly decayed following:
\begin{equation}
    \label{eq:epsilon_greedy}
    \epsilon_i = \max \left( \epsilon_{i,\text{min}}, 1 - \frac{(1 - \epsilon_{i,\text{min}}) \cdot n}{\epsilon_{i,\text{frac}} \cdot N} \right)
\end{equation}
with $\epsilon_{\rm frac}$ representing the fraction of episodes over which the decay occurs, $\epsilon_{\rm min}$ the minimum exploration probability the agent will maintain after the decaying phase, and $n$ the current episode. 
This exploration policy is a well-known approach, adopted widely in the literature due to its simplicity. 
However, it can lead to suboptimal performance or slower training time due to the complex tuning of the $\varepsilon_{i, \rm frac}$ hyperparameter.
While other exploration alternatives exist, we focus on $\varepsilon$-greedy as a standard baseline to enabling more direct comparison with the proposed meta-advisor. Additionally, several alternatives rely on state visitation or confidence estimates whose applicability under partial observability and multi-agent non-stationarity can be limited, whereas $\varepsilon$-greedy remains simple and broadly applicable.
A schematic representation of applying $\varepsilon$-greedy exploration to a generic set of tasks $\mathcal{T}$ is show in Fig.~\ref{fig:epsilon-greedy}

\subsection{Multi-Agent Meta-Advisor}
The random action selection of the $\epsilon$-greedy approach can negatively affect navigation by causing unnecessary
exploration that reduces learning efficiency.
Moreover, it lacks mechanisms to exploit similarities among tasks $\tau_i$ to make the exploration phase more efficient. These limitations reduce the time-to-proficiency of the fleet, critical for the online learning system adopted.
To overcome this, we introduce a context-aided deep multi-agent meta-advisor that guides the exploration phase using prior knowledge from the whole set of tasks $\mathcal{T}$.
This advisor learns a shared exploration policy $\mu$ that provides exploration actions $\Tilde{a}_t^{(u)}$ to agents based on an augmented observation $\Tilde{o}_t^{(u)}= o_t^{(u)} \mathbin\Vert \text{Id}(\omega_{\tau_i})$, with $o_t^{(u)}$ the original agent observation and $\text{Id}()$ a function mapping the service area $\omega_{\tau_i}$ to a one-hot encoded vector.
This additional context enhances the robustness of the advisor to heterogeneous service areas while maintaining generalization across all tasks in $\mathcal{T}$, as previously studied in our prior work~\cite{11011207}.
Specifically, the advisor selects the exploration action $\Tilde{a}_t^{(u)}$ by maximizing its action-value function:
\begin{equation}
    \tilde{a}_t^{(u)} = \arg\max\limits_{a} Q_{\mu}(\tilde{o}_t^{(u)}, a | \theta_\mu),
\end{equation}
with $\theta_\mu$ the parameters of the meta-advisor network.

The training of the advisor happens in parallel with the individual agents: while each agent learns its task-specific exploitation policy $\pi_{\tau_i}$, the meta-advisor learns $\mu$ from the collective experiences obtained by agents in all the tasks and stored in a shared replay buffer $\mathcal{K}_\mu$. The existing core network supports the exchange of experiences among tasks, eliminating the need for additional wireless transmissions.
Without loss of generality, the model is trained via the \ac{3DQN} algorithm with parameters $\theta_\mu$ and $\theta_\mu^-$, and an $\varepsilon_\mu$-greedy policy assuming $\varepsilon_{\mu,\rm frac} << 1$ for its training. We adopt this value-based off-policy formulation to exploit experience replay and improve sample-efficiency of the advisor across tasks; policy-based meta-advisors are possible, but are left for future work.
When the meta-advisor is employed as exploration policy, $\varepsilon_i$ refers to the probability of selecting an exploration action for agent $i$ for task $\tau_i$, which can either random or selected by the meta-advisor depending on the value of $\varepsilon_\mu$.
It is highlighted that the advisor acts as an external module overseeing the parallel training of other agents, thus it can be trained with different hyperparameters or \ac{DRL} algorithms.
A representation of the meta-advisor architecture is reported in Fig.~\ref{fig:MAMA}.

\subsection{Advisor Override}
%
%
%
%
In our approach, the combination of a task-specific model, a meta-advisor, and a generic exploration policy provides each agent with three distinct types of candidate actions at each time step, namely:
\begin{itemize}
    \item \textit{exploit}, where the agent chooses the best action exploiting the current task-specific model learned, $\overline{a}_t^{(u)}$;
    \item \textit{meta}, where the agent uses the action provided by the trained meta-advisor to perform safe and improved exploration, $\Tilde{a}_t^{(u)}$;
    \item \textit{random}, where the agent performs standard random exploration. 
\end{itemize}
The process of choosing which of these three action types to execute is what we define as an agent specific \textit{behavior selection}. To formalize this, we define the preliminary behavior selection variable $\Tilde{\zeta}_t^{(u)}$:
\begin{equation}
     \Tilde{\zeta_t}^{(u)} =   
        \begin{cases}
        \text{random} & \text{with prob. } \epsilon_i\cdot\epsilon_\mu \\
        \text{meta} & \text{with prob. } \epsilon_i\cdot(1-\epsilon_\mu) \\
        \text{exploit} & \text{with prob. } (1-\epsilon_i)\\
        \end{cases}
\end{equation} 

This selection of behaviors is random and based on the current values of $\varepsilon_i$ and $\varepsilon_\mu$, which depend on the hyperparameters set at the beginning of training. 
%
However, the potential overconfidence of the meta-advisor in prescribing a specific meta exploration action $\tilde{a}$ poses a risk of constraining agents, particularly when encountering novel states within a given task.
Furthermore, the original formulation may not fully account for the inherent non-stationarity of multi agent and multi task environments. 
In such settings, the shared advisor model may inherently bias its convergence towards a subset of simpler tasks, where rewards are more readily accessible, while failing to provide adequate guidance for more complex scenarios.
This imbalance could lead to a biased advisor that disproportionately benefits certain agents or tasks at the expense of others.
This can be more evident in the case of fleet operations, where each task requires discovering a specific cooperative trajectory. 
To this end, we introduce the concept of \textit{advisor override}. 
This technique allows an agent with a preliminary mode $\rm meta$ at a given time step, i.e.  $\tilde \zeta_t^{(u)} = \mathrm{meta}$, to switch to random behavior mode if this condition is met:
\begin{equation}
    \begin{split}
        & \mathcal{C}^{(u)}_{t,\mathrm{m}\to\mathrm{r}}  \triangleq 
         \;\big[ \tilde \zeta_t^{(u)} = \mathrm{meta} \big] \wedge {} \\
         & \qquad \left( Q_{\pi_i}\!\left(o_t^{(u)}, \tilde{a}_t^{(u)} \, \vert \,\theta_i \right) < \frac{1}{\vert \mathcal{A} \vert} \sum_{a\in\mathcal{A}} Q_{\pi_i}\!\left( o_t^{(u)},a \, \vert \, \theta_i\right) \right)
    \end{split}
    \label{eq:override_condition}
\end{equation}
where, $o_t^{(u)}$ is the current agent observation, $\tilde{a}_t^{(u)}$ is the action provided by the meta-advisor, and $Q_{\pi_i}(o_t^{(u)},a \vert \, \theta_i)$ are the Q-value calculated for a generic action $a$ and the current observation given the task-specific policy $\pi_i$.
In other words, the advisor override is triggered whenever the task-specific model for task $\tau_i$ predicts that the action suggested by the meta-advisor yields a lower long-term reward, measured by its estimated Q-value, than the average reward expected from the other available actions.
This mechanism has twofold advantages. On one hand, it allows the advisor to gather more heterogeneous experiences, avoiding overconfidence and overfitting in the early stage of training. On the other hand, it protects the exploitation model $\pi_i$ from a biased meta-advisor, allowing it to perform standard exploration when necessary. 
It is important to notice that in the case of advisor rejection, the agent will perform standard random exploration considering the whole action space, with a non-zero probability of selecting the rejected action.
This allows for ignoring the contribution of the advisor while minimizing the potential biases introduced by this mechanism.
Advisor override draws inspiration from~\cite{ganapathi2022multi}, however, it does not require a pre-training phase, making this technique feasible for online and parallel training among multiple tasks. 
The final behavior mode is then derived as:
\begin{equation}
\zeta_t^{(u)} \;=\;
\begin{cases}
\mathrm{random}, & \text{if }\mathcal{C}^{(u)}_{t,\mathrm{m}\to\mathrm{r}}\ \text{holds}\\[4pt]
\tilde \zeta_t^{(u)}, & \text{otherwise}.
\end{cases}
\end{equation}
The overall \ac{MAMO} algorithm is reported in Algorithm~\ref{alg:main_mamo}.
\input{Algo/MAMO}

%% file: Algo/MAMO.tex
\begin{algorithm}[t]
\centering
\caption{MAMO online training}\label{alg:main_mamo}
\footnotesize
\begin{minipage}{0.9\columnwidth}
\begin{algorithmic}[1]
\STATE Initialize advisor network $\theta_\mu$, initialize task networks $\theta_i, i=1, \dots ,\vert \mathcal{T} \vert$ 
\FOR{$n=0, \dots, N-1$}
    \FOR{$\tau_i \in \mathcal{T}$}
        \STATE $s_0 \gets {s}_{0,\tau_i}$
        \STATE update $\epsilon_i$ according to \eqref{eq:epsilon_greedy} with $\epsilon_{\rm frac}=\epsilon_{i,\rm frac}$
        \STATE update $\epsilon_\mu$ according to \eqref{eq:epsilon_greedy} with $\epsilon_{\rm frac}=\epsilon_{\mu,\rm frac}$
        \FOR{$t=0, \dots, T-1$}
            \FOR{$u\in\mathcal{U}$}
                \STATE $\Tilde{o}^{(u)}_t \gets {o^{(u)}_t}\mathbin\Vert \text{Id}(\omega_{\tau_i})$
                \STATE $\tilde{\zeta}_t^{(u)} \gets
                    \begin{cases}
                    \mathrm{random}, & \text{with prb. } \epsilon_i\cdot\epsilon_\mu \\
                    \mathrm{meta}, & \text{with prb. } (1-\epsilon_\mu)\cdot\epsilon_i \\
                    \mathrm{exploit}, & \text{otherwise}
                    \end{cases}$
                \vspace{1ex}
                \STATE $\zeta_t^{(u)} \gets
                    \begin{cases}
                    \mathrm{random}, & \begin{tabular}{@{}l@{}} 
                                        if adv. override is enabled \\ 
                                        and $\mathcal{C}^{(u)}_{t,\mathrm{m}\to\mathrm{r}}$ holds 
                                       \end{tabular} \\[3ex]
                    \tilde{\zeta}_t^{(u)}, & \text{otherwise}
                    \end{cases}$
                \vspace{1ex}
                \STATE $a_t^{(u)} \gets
                    \begin{cases}
                    \text{random}, & \hspace{-0.8em} \zeta_t^{(u)} = \mathrm{random} \\[1ex]
                    \arg\max\limits_{a} Q_{\mu}(\Tilde{o}^{(u)}_t, a), & \hspace{-0.8em} \zeta_t^{(u)} = \mathrm{meta} \\[1ex]
                    \arg\max\limits_{a} Q_{\pi_i}(o_t^{(u)}, a), & \hspace{-0.8em} \zeta_t^{(u)} = \mathrm{exploit}
                    \end{cases}$
            \ENDFOR
            \STATE $s_t\gets \{ o_t^{(u)}\}_{u\in\mathcal{U}}$
            \STATE $a_t\gets \{ a_t^{(u)}\}_{u\in\mathcal{U}}$
            \STATE $r_t \sim R_i(s_t,a_t)$
            \STATE $s_{t+1}\sim P_i(s_t,a_t)$
            \FOR{$u\in\mathcal{U}$}
                \STATE $\Tilde{o}^{(u)}_{t+1} \gets 
                    {o^{(u)}}_{t+1}\mathbin\Vert \text{Id}(\omega_{\tau_i})$
                \STATE $\mathcal{K}_i \gets \{ o^{(u)}_t, a^{(u)}_t, r^{(u)}_t, o^{(u)}_{t+1}\}$
                \STATE $\mathcal{K}_\mu \gets \{ \Tilde{o}^{(u)}_t, a^{(u)}_t, r^{(u)}_t, \Tilde{o}^{(u)}_{t+1}\}$ 
            \ENDFOR
            \STATE Train $\theta_i$ with batch $k \sim \mathcal{K}_i$ according to \eqref{eq:loss}
        \ENDFOR
    \ENDFOR
    \FOR{$j=0, \dots, J-1$} 
        \STATE Train $\theta_\mu$ with batch $k \sim \mathcal{K}_\mu$ according to \eqref{eq:loss}
    \ENDFOR
\ENDFOR
\end{algorithmic}
\end{minipage}
\end{algorithm}

%% file: Chapters/Numerical_Results.tex
\section{Numerical Results}
\label{sec:results}

\begin{figure*}[t]
\centering
\subfloat[$\omega_1$\label{fig:map_saffi}]{%
    \frame{\includegraphics[width=.30\linewidth, trim={0 0 0 0}, clip]{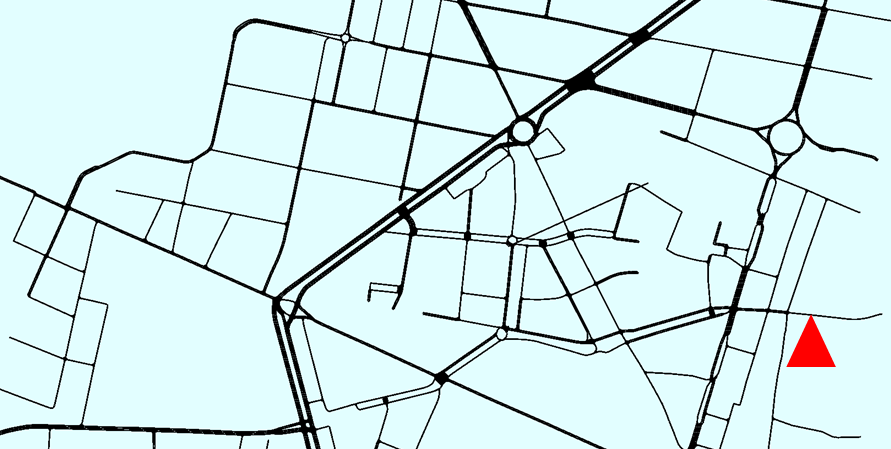}}
}
\hfill
\subfloat[$\omega_2$ \label{fig:map_pertini}]{%
    \frame{\includegraphics[width=.30\linewidth, trim={0 0 0 0}, clip ]{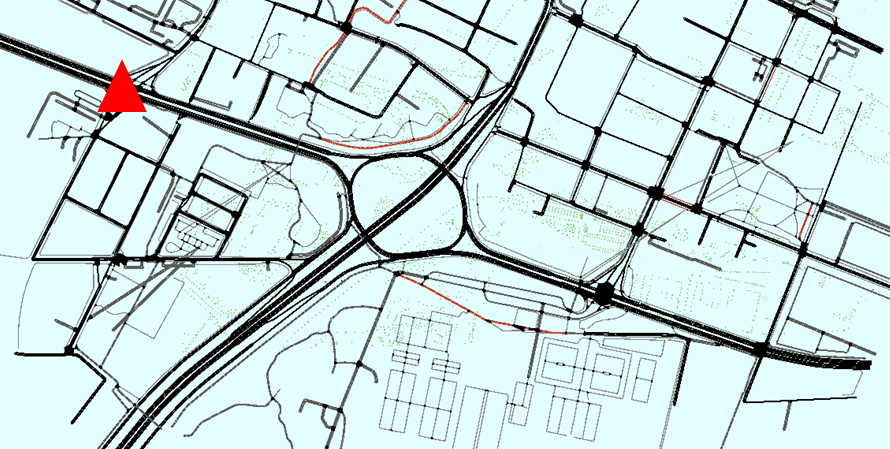}}
}
\hfill
\subfloat[$\omega_3$ \label{fig:map_saragozza}]{%
    \frame{\includegraphics[width=.30\linewidth, trim={0 0 0 0}, clip]{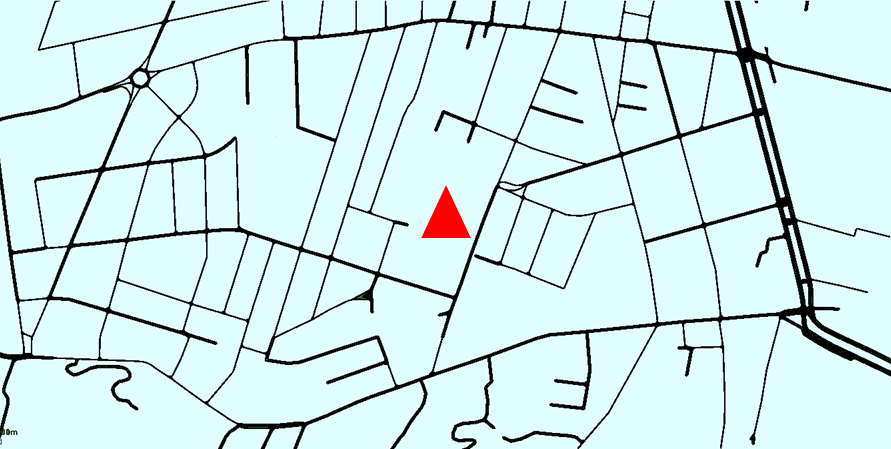}}
}
\caption{Service area considered, with red triangles representing the \ac{MBS} position.}
\label{fig:services_map}
\end{figure*}

\subsection{Simulation Settings}
This section details the evaluation of the proposed \ac{MAMO} exploration policy for multi-scenario fleet trajectory design.
We consider a service area set $\Omega=\{\omega_1, \omega_2, \omega_3\}$, modeled after real districts in Bologna, Italy.
These regions share the same dimensions $L \times W$,
but differ in street layout map and in the number \acp{GUE} traversing the area. 
Within each area, 
a \ac{UABS} fleet $\mathcal{U}$ ($\vert \mathcal{U} \vert = 3$) is deployed to enhance \ac{GUE} connectivity and service continuity in coordination with terrestrial \acp{MBS}.
Fig.~\ref{fig:services_map} illustrates the service areas, detailing the specific street layout map, and \ac{MBS} position (depicted as a red triangle) for each. \ac{GUE} mobility is simulated using \ac{SUMO}~\cite{SUMO2018} to ensure realistic micro-mobility behavior under given road traffic distributions.

The whole task set $\mathcal{T}$ consists of combinations of service areas and possible \ac{UABS} fleet takeoff positions. Specifically, we define five possible takeoff settings: 
$\{[\xi_1,\xi_2,\xi_4]$, $[\xi_1,\xi_3,\xi_4]$, $[\xi_1,\xi_4,\xi_5]$, $[\xi_2,\xi_3,\xi_5]$, $[\xi_6, \xi_7, \xi_8]\}$,
where $\xi_{k}=(x_k, y_k)$ represents predefined takeoff sites in each region.
Tab.~\ref{tab:sim_param} summarizes the simulation parameters.

For each task, agents are trained for $N$ episodes using the following learning strategies for comparison:
\begin{itemize}
    \item \textbf{\ac{MAMO}:} The proposed approach, where agents train a task-specific models $\theta_i$ exploiting the multi-agent meta advisor as an exploration policy, with advisor override enabled. 
    \item \textbf{MAMA:} Agents train task-specific models $\theta_i$ using the multi-agent meta-advisor, but the advisor override mechanism is disabled, representing an ablation of the proposed solution.
    \item \textbf{$\varepsilon$-greedy:} Standard training using independent exploration between tasks, tested with varying decay rates $\varepsilon_{i,\rm frac}$.
    \item \textbf{Generalized:} A single model is trained for the entire task set $\mathcal{T}$. All agents share the same model for trajectory planning without per-task specialization.
\end{itemize}
For \ac{MAMO} and MAMA, the meta-advisor model was trained in parallel with the task-specific models. 

\begin{table*}[t]
\centering
\footnotesize
\begin{tabularx}{\linewidth}{| Y | c | Y | c | Y | c || c | Y |}
\hline
$ v^{(\rm u)} $ & 20 m/s & $h^{(\rm u)}$ & 100 m & $\phi$ & 100° & $\xi_1$ & $(0, 0)$ \\
\hline
$T$ & 270 s & $T_s$ & 10 s & $|\mathcal{G}|$ $[\omega_1, \omega_2, \omega_3]$ & $[200, 180, 90]$ & $\xi_2$ & $(L, 0)$ \\
\hline
$f_c$ & 30 GHz & $\B$ & 9 & $P_{\rm tx}$ & 14 dBm & $\xi_3$ & $(0, W)$ \\
\hline
$G_{\rm tx}$ & 0 dB & $G_{\rm rx}$ & 23 dB & $P_{\rm n}$ & -106 dBm & $\xi_4$ & $(L, W)$ \\
\hline
$\epsilon_{\mu,{\rm frac}}$ & 0.2 & $\epsilon_{i,{\rm frac}}$ & 0.6 & $\epsilon_{\rm min}$ & 0.05 & $\xi_5$ & $(\tfrac{L}{2}, \tfrac{W}{2})$ \\
\hline
$\vert \mathcal{K}_i \vert$ & 5e+4 & $\vert \mathcal{K}_\mu \vert$ & 1e+6 & $k$ & 128 & $\xi_6$ & $(\tfrac{L}{4}, \tfrac{3W}{4})$ \\
\hline
$Y$ & 100 & $\gamma$ & 0.99 & $J$ & 2700 & $\xi_7$ & $(\tfrac{3L}{4}, \tfrac{W}{4})$ \\
\hline
$D$ & 1 MBit & $L$ & 1500 m & $W$ & 700 m & $\xi_8$ & $(\tfrac{3L}{4}, \tfrac{3W}{4})$ \\
\hline
\end{tabularx}
\caption{Simulation parameters.}
\label{tab:sim_param}
\end{table*}

\subsection{Performance Metrics}
We evaluate performance based on the average return over a task set: 
\begin{equation}
    \overline{R}(n) = \frac{1}{\vert \T \vert} \sum_{\tau_i \in \T} R_{\tau_i,n}
\end{equation}
where $R_{\tau_i,n}=\left( \sum_{u\in\mathcal{U}} \sum_{t=0}^T r_t^{(u)} \right)$ represents the total reward obtained by the fleet solving task $\tau_i$ at the $n$-th training episode. For clarity, results are smoothed using a moving average with a window length of 50.

To highlight the efficiency of the proposed meta-exploration approach, we quantify the time-to-proficiency concept via the \textit{\ac{FSE}} metric, denoted by $\widehat{n}_{\tau_i}$. The \ac{FSE} is the earliest episode in the training sequence of task $\tau_i$ where the return $R_{\tau_i,n}$ exceeds a success threshold $R_{\rm th}$:
\begin{equation}
    \widehat{n}_{\tau_i} = \arg\min_n \left\{n \hspace{.5em} | \hspace{.5em} {R_{\tau_i,n}} \geq R_\text{th}\right\}
\end{equation}
Reported values are averaged across multiple independent simulation runs for each approach.

We also track the episode \textit{\ac{AOC}}, defined as:
\begin{equation}
    \text{AOC}_{\tau_i}(n) = \sum_{t=0}^T \sum_{u\in\mathcal{U}} {I}_{\mathcal{C}^{(u)}_{t,\rm{m}\rightarrow \rm{r}}}
\end{equation}
where ${I}_{\mathcal{C}^{(u)}_{t,\rm{m}\rightarrow \rm{r}}}$ is an indicator function equal to $1$ when the override condition $\mathcal{C}^{(u)}_{t,\rm{m}\rightarrow \rm{r}}$ holds, and $0$ otherwise. This metric quantifies how frequently the advisor override is triggered during episode $n$ of task $\tau_i$.

Finally, to assess the continuity of service provided by the \ac{UABS} fleet, we report the \textit{percentage of satisfied users} $P_g$, calculated as:
\begin{equation}
    P_g = \frac{1}{\vert \G \vert} \sum_{g \in \G} \frac{N_{g}^{(sat)}}{N_g}
\end{equation}
where $N_g$ is the total number of service windows for the $g$-th \ac{GUE}, and $N_{g}^{(\rm sat)}$ denotes the windows where the service constraint $N_s \ge \hat{N_s}$ is met.

\subsection{Training Performance Analysis}
\begin{figure*}
    \centering
    \footnotesize
    \includegraphics[width=1\linewidth ,trim={5 11 8 12}, clip ]{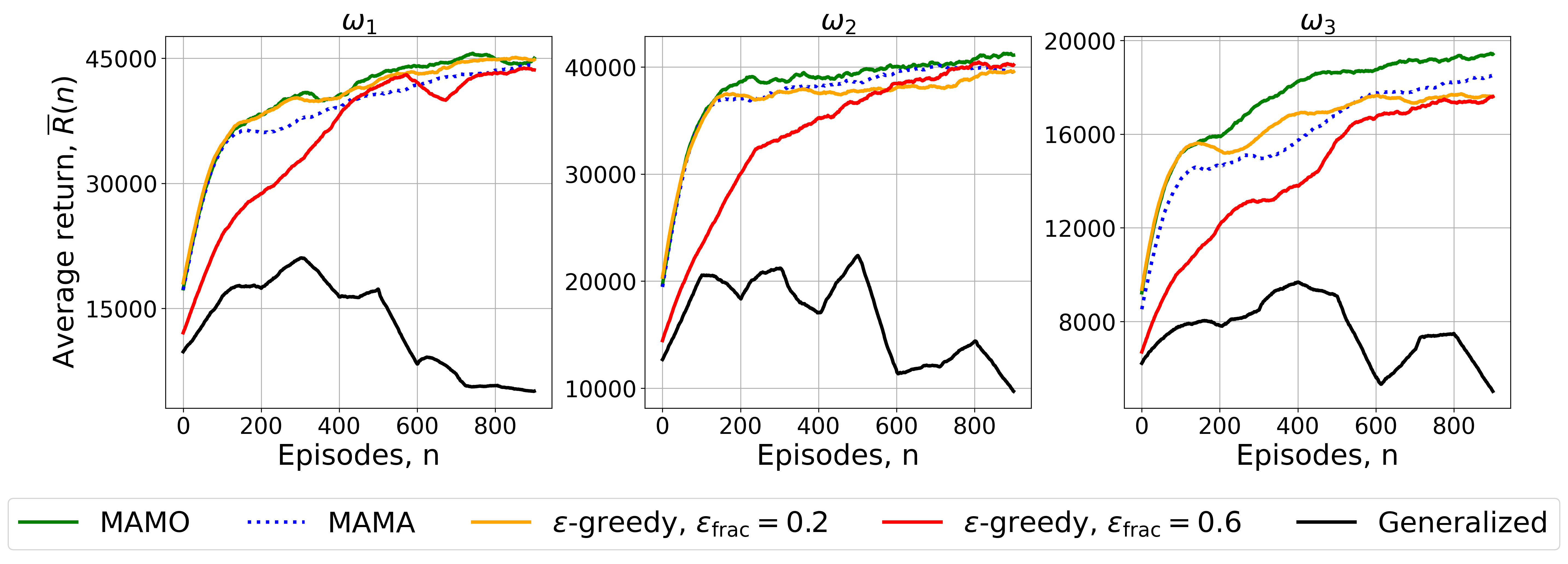}
    \caption{Average return trends during training, averaged over tasks belonging to the same service area.}
    \label{fig:MA_training}
\end{figure*}
\begin{figure*}
    \centering
    \includegraphics[width=1\linewidth]{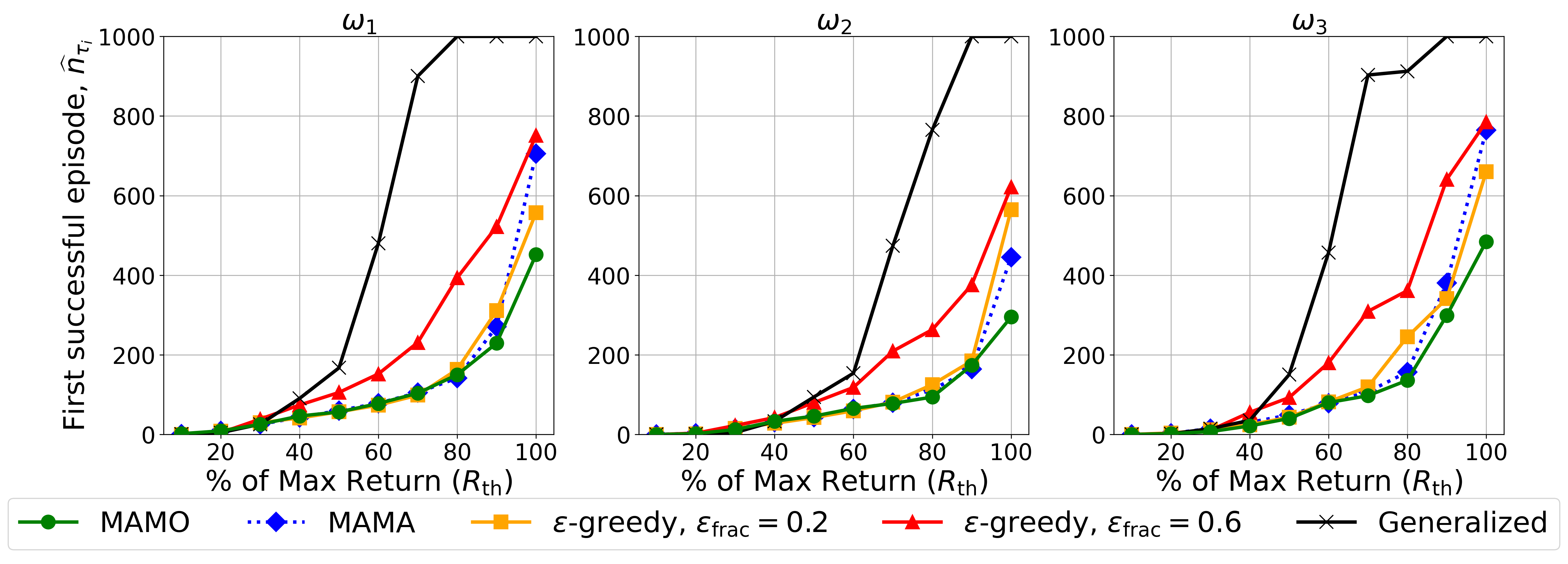}
    \footnotesize
    \caption{First Successful Episode ($\widehat{n}_{\tau_i}$) as a function of the target return threshold, expressed as a percentage of the maximum return obtained during training. A lower curve indicates superior performance with the target reached in fewer episodes.}
    \label{fig:FSE}
\end{figure*}
\begin{figure}
    \centering
    \includegraphics[width=\linewidth,trim={0 0 0 0}, clip]{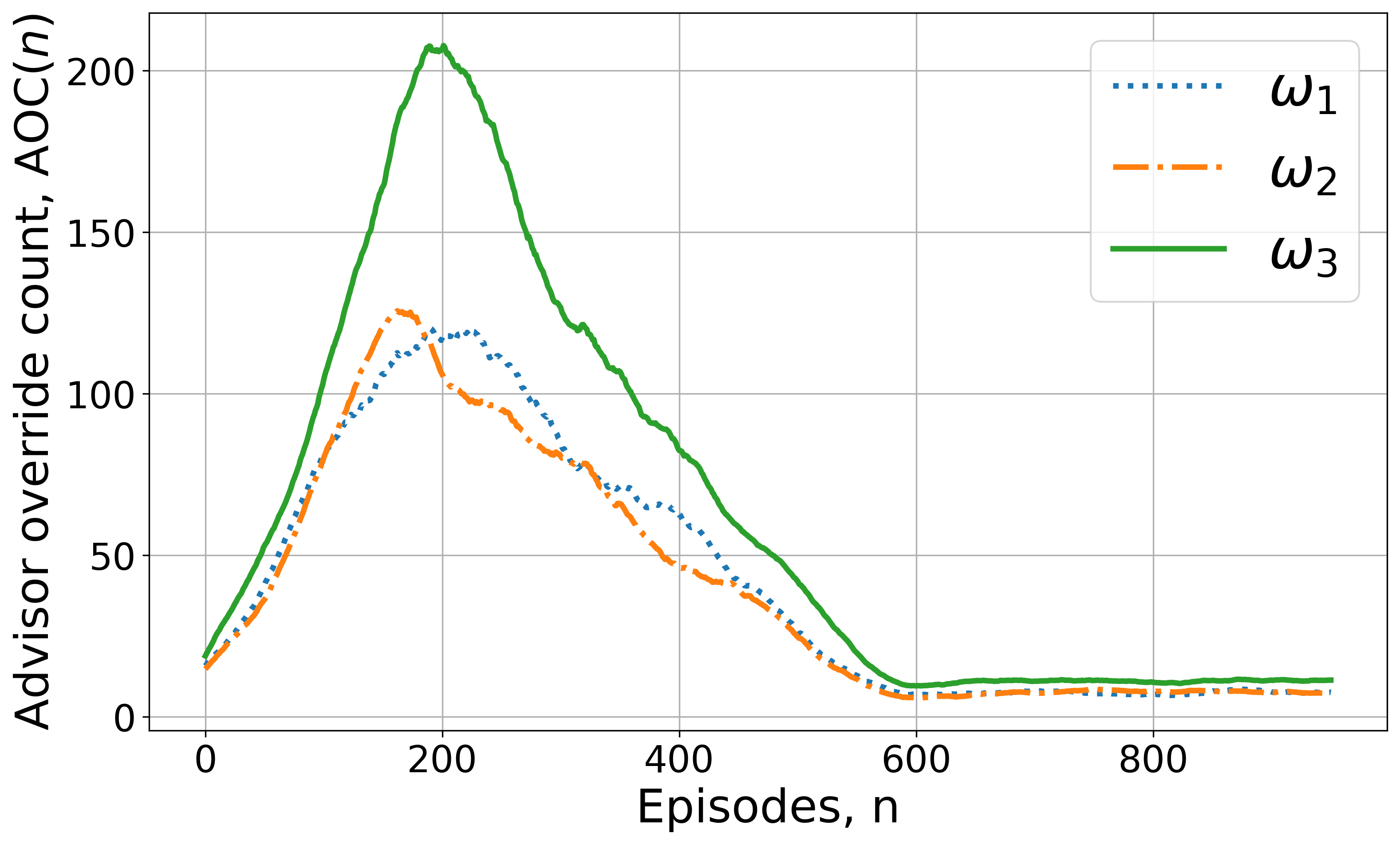}
    \caption{Evolution of the Advisor Override Count (AOC) during training for different service areas.}
    \label{fig:adv_count}
\end{figure}
Fig.~\ref{fig:MA_training} illustrates the cumulative return averaged over all tasks associated with a specific service area. Note that while the training process considers the entire task set $\mathcal{T}$ concurrently, the results are grouped by service area for clearer analysis, as each region is characterized by a different number of users, leading to different reward magnitudes.
The proposed \ac{MAMO} exploration policy consistently outperforms the benchmarks, demonstrating superior convergence speed and higher final returns. This robustness is maintained across all service areas despite significant variations in reward scale and environmental layout.
A critical observation arises from the comparison with MAMA in the service area $\omega_3$. Here, the MAMA approach, lacking the advisor override mechanism, suffers a significant performance degradation compared to \ac{MAMO}. This suggests that in specific scenarios, the meta-advisor may provide guidance that is less effective for a given task due to overfitting toward other tasks. Without the ability to reject these suggestions via the override mechanism, agents may follow less effective advice, limiting the approach's generalization capabilities and exploration proficiency. 
Regarding the standard $\varepsilon$-greedy baselines, results highlight the trade-off inherent in hyperparameter tuning. A slow decay ($\varepsilon_{\rm frac}=0.6$) delays convergence significantly. Conversely, a fast decay ($\varepsilon_{\rm frac}=0.2$) accelerates initial learning but leads to premature convergence to sub-optimal policies and plateauing, as evident in $\omega_2$ and $\omega_3$, where it fails to match the final performance of \ac{MAMO}.
Finally, the \textit{Generalized} approach fails to converge to a viable policy, underscoring the necessity of task-specific models; thus, a single shared policy cannot capture the distinct fleet trajectory requirements of different urban environments.

\input{Table/win_ratio_vertical}

While the performance trends in Fig.~\ref{fig:MA_training} highlight the superiority of \ac{MAMO}, a more granular comparative analysis of the return for each task $\tau_i$ is provided in Tab.~\ref{tab:win_ratios_vertical} via the \textit{Win-Ratio}.
It represents the percentage of training episodes $n$, out of the total $N$, in which the return $R_{\tau_i,n}$ achieved by \ac{MAMO} is higher than that obtained by the considered benchmark at the same training episode $n$.
Therefore, this metric captures the dominance of \ac{MAMO} throughout the entire learning phase, with lower values indicating a more challenging opponent.
According to the results in Tab.~\ref{tab:win_ratios_vertical}, MAMA and $\varepsilon$-greedy with $\varepsilon_{\rm frac}=0.2$ appear as the most competitive on average. However, the leaderboard varies significantly across tasks, confirming that classical and naive advisor exploration approaches lack robustness against task variations, whereas \ac{MAMO} maintains consistent superiority.

Fig.~\ref{fig:FSE} analyzes the training efficiency by plotting the \ac{FSE}. We track the number of episodes required to reach a specific percentage of the achieved maximum return per task. 
For this metric, a curve with slower growth indicates that high-performance thresholds are reached earlier in the training process.
Across all areas, \ac{MAMO} exhibits the flattest learning curves.  In contrast, benchmark solutions, particularly Generalized and $\varepsilon$-greedy, show an asymptotic vertical rise near the 90\%-100\% return mark, indicating a struggle to refine the policy towards optimality within the 1000 training episodes. This capability to reach near-optimal returns in a fraction of the episodes is critical for online and safe learning systems, where reducing the ``time-to-proficiency'' directly translates to an immediate boost in network performance.

To explain the performance gap observed in Fig.~\ref{fig:MA_training}, Fig.~\ref{fig:adv_count} details the \ac{AOC} trends. The override frequency generally peaks between episodes 150 and 250, corresponding to the transition from random to meta-guided exploration.
Most notably, the override count for service area $\omega_3$ reaches a peak of $\approx 210$, nearly double the count observed for $\omega_1$ and $\omega_2$. This high rejection rate indicates that the advisor is less reliable in the $\omega_3$ scenario. This correlates perfectly with the failure of MAMA in Fig.~\ref{fig:MA_training}. MAMA agents are forced to follow these task-mismatched suggestions, while \ac{MAMO} agents trigger the protection mechanism, reverting to random exploration to avoid local optima.

\begin{figure}
    \centering
    \includegraphics[width=0.75\linewidth,trim={0 0 0 0}, clip]{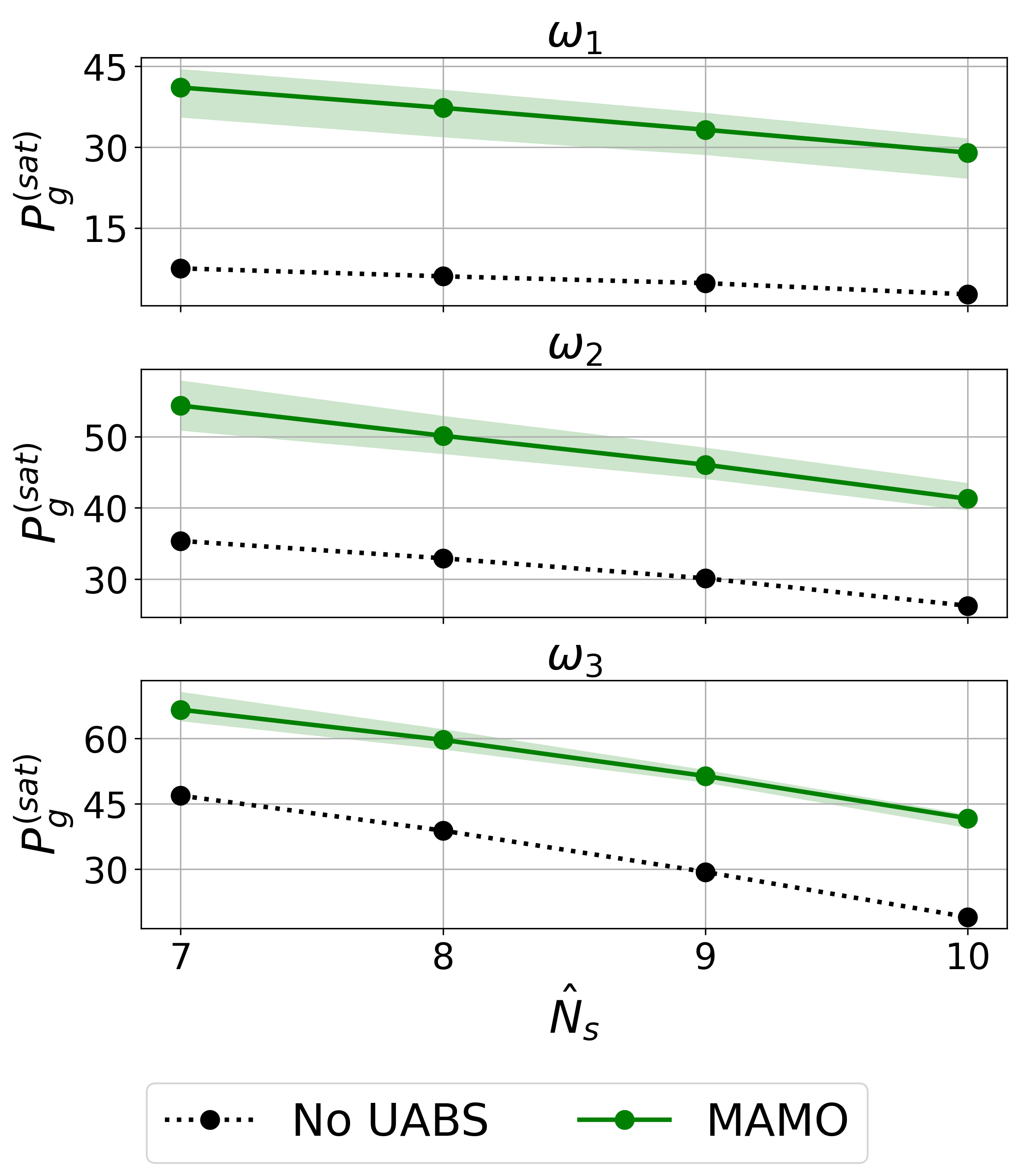}
    \caption{Percentage of satisfied users ($P_g$) for the \ac{MAMO}-trained fleet as a function of the service window threshold $\hat{N}_s$. The solid line represents the average across tasks, while the shaded area shows the minimum and maximum.}
    \label{fig:pg}
\end{figure}

\begin{figure}[t]
\centering
\subfloat[$t$=0\label{fig:traj_saragozza_0}]{%
    \frame{\includegraphics[width=.8\linewidth, trim={10 10 10 10}, clip]{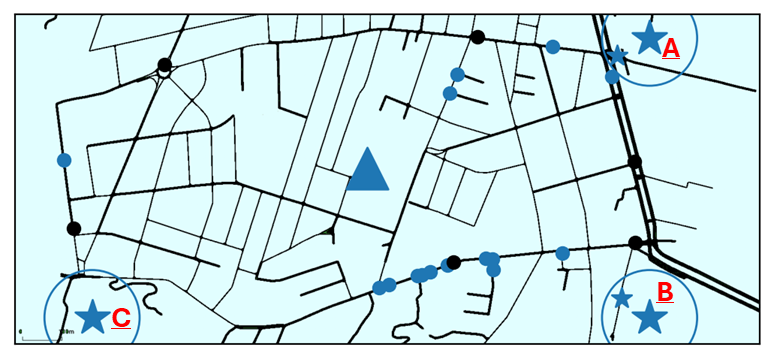}}
}
\hfill
\subfloat[$t$=90 \label{fig:traj_saragozza_1}]{%
    \frame{\includegraphics[width=.8\linewidth, trim={10 10 10 10}, clip ]{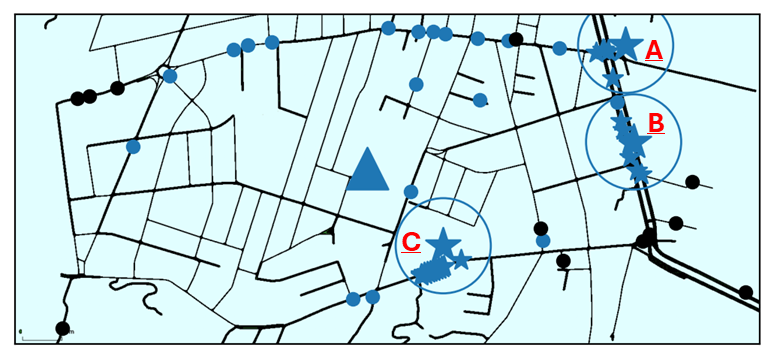}}
}
\hfill
\subfloat[$t$=230 \label{fig:traj_saragozza_2}]{%
    \frame{\includegraphics[width=.8\linewidth, trim={10 10 10 10}, clip]{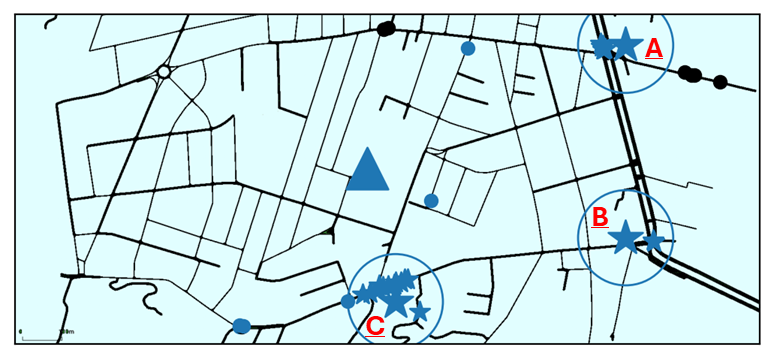}}
}
\caption{Example of fleet trajectory learned by the \acp{UABS}. Star dots represent \ac{UABS}, with circles their covered area. The triangle represents the \ac{MBS}. (a) $t$=0, initial takeoff position (b) $t$=90 flight mission, \ac{UABS} A serves users at the traffic light in upper-right corner, with \ac{UABS} B close enough to ensure service continuity. \ac{UABS} C focuses on users in the lower part of the service area (c) $t$=230, \ac{UABS} B has moved to follow the available user, while \ac{UABS} C changed its position to serve a greater amount of users.}
\label{fig:traj_snapshot}
\end{figure}


\subsection{Network Performance Analysis}
\input{Table/load_distribution_table_auto}
In this section, we evaluate the learned adaptive trajectory model from a network performance perspective, focusing on agents trained via \ac{MAMO}.
Tab.~\ref{tab:load_distribution} summarizes the network load distribution between the \ac{MBS} and each \ac{UABS} together with the total amount of received packets. These results represent averages across tasks within the same service area over episodes of duration $T$. 
These results indicate that the integration of the aerial segment significantly enhances overall network throughput even when the \ac{MBS} and \ac{UABS} fleet share the same system bandwidth. This improvement is most evident in service area $\omega_1$. Here the \ac{MBS}, limited by its fixed location and \ac{NLoS} dominated channel conditions, fails to meet \acp{GUE} uplink demand. In contrast, the deployment of the fleet increases the amount of received packets by more than 200\%. Moreover, the load distribution among \acp{UABS} is balanced. This demonstrates the effective cooperative behavior that emerges within the fleet through the proposed \ac{MADRL} learning framework.

Regarding the continuity of service, Fig.~\ref{fig:pg} reports the percentage of satisfied users, $P_g$, as a function of different service thresholds $\hat{N}_s$, averaged among tasks (solid line) and within the minimum and maximum reached in different tasks. 
Performance is heavily dependent on the specific task characteristics, as final network performance depends on both the \ac{UABS} fleet trajectory design and the service provided by the \ac{MBS}.
Interestingly, despite being the most challenging for the advisor (as seen in the \ac{AOC} analysis), service area $\omega_3$ allows for the highest quality of service, with satisfied user rates reaching up to 70\%. In contrast, $\omega_1$ presents a stricter environment, capping at approximately 45\%. In all scenarios, the \ac{MAMO}-trained fleet maintains service continuity, although satisfaction naturally decreases as the service constraint $\hat{N}_s$ increases.
Finally, the contribution provided by the \ac{UABS} fleet considerably improves the user \ac{QoE} when compared to a scenario where only the \ac{MBS} is present, as suggested by the gap between the solid and dotted curves.

Finally, Fig. \ref{fig:traj_snapshot} shows an example of the evolution of the fleet trajectory considering a generic task within service area $\omega_3$. Throughout a flight mission, \acp{UABS} intelligently reposition themselves to provide continuity of service to \acp{GUE}.

%% file: Table/win_ratio_vertical.tex
\begin{table}[t]
\centering
\caption{Win Ratio of proposed \ac{MAMO} vs benchmarks. Bold results represent the strongest competitor for the task considered.}
\resizebox{\columnwidth}{!}
{
\begin{tabular}{l||cccc}
\toprule
\textbf{Task} & \textbf{MAMA} & $\varepsilon$-greedy 0.2 & $\varepsilon$-greedy 0.6 & \textbf{Gen}\\
\midrule
\multicolumn{5}{l}{$\omega_1$} \\
$[\xi_1,\xi_2,\xi_4]$ & 41.3\% & \textbf{3.4}\% & 45.6\% & 100.0\% \\
$[\xi_1,\xi_3,\xi_4]$ & 73.6\% & \textbf{63.1}\% & 81.6\% & 100.0\% \\
$[\xi_1,\xi_4,\xi_5]$ & 95.6\% & \textbf{95.5}\% & 97.0\% & 100.0\% \\
$[\xi_2,\xi_3,\xi_5]$ & 61.2\% & \textbf{21.9}\% & 98.9\% & 100.0\% \\
$[\xi_6, \xi_7, \xi_8]$ & 89.8\% & \textbf{44.4}\% & 99.9\% & 100.0\% \\
\midrule
\multicolumn{5}{l}{$\omega_2$} \\
$[\xi_1,\xi_2,\xi_4]$ & 73.7\% & \textbf{68.9}\% & 79.8\% & 100.0\% \\
$[\xi_1,\xi_3,\xi_4]$ & \textbf{67.8}\% & 79.6\% & 85.0\% & 100.0\% \\
$[\xi_1,\xi_4,\xi_5]$ & 92.6\% & \textbf{77.6}\% & 87.9\% & 100.0\% \\
$[\xi_2,\xi_3,\xi_5]$ & \textbf{59.7}\% & 65.6\% & 90.4\% & 100.0\% \\
$[\xi_6, \xi_7, \xi_8]$ & \textbf{49.4}\% & 80.3\% & 95.8\% & 100.0\% \\
\midrule
\multicolumn{5}{l}{$\omega_3$} \\
$[\xi_1,\xi_2,\xi_4]$ & \textbf{56.0} \% & 85.8\% & 100.0\% & 100.0\% \\
$[\xi_1,\xi_3,\xi_4]$ & 86.6\% & \textbf{83.7}\% & 98.9\% & 100.0\% \\
$[\xi_1,\xi_4,\xi_5]$ & 96.4\% & \textbf{68.9}\% & 100.0\% & 100.0\% \\
$[\xi_2,\xi_3,\xi_5]$ & 100.0\% & \textbf{94.4}\% & 100.0\% & 100.0\% \\
$[\xi_6, \xi_7, \xi_8]$ & 98.1\% & \textbf{68.3}\% & 70.3\% & 100.0\% \\
\midrule
Average & 76.1\% & \textbf{66.8}\% & 88.7\% & 100.0\% \\
\bottomrule
\end{tabular}%
}
\label{tab:win_ratios_vertical}
\end{table}

%% file: Table/load_distribution_table_auto.tex
%
\begin{table}[t]
\centering
\caption{Traffic Load Distribution: MBS vs. Drone Fleet}
\resizebox{\columnwidth}{!}{%
\begin{tabular}{l||ccccc}
\toprule
\textbf{Scenario} &  \textbf{Total Load} & \textbf{MBS Load} & \textbf{UABS A} & \textbf{UABS B} & \textbf{UABS C} \\
 & \textbf{[Rx Packet]} & \textbf{[\%]} & \textbf{[\%]} & \textbf{[\%]} & \textbf{[\%]} \\
\midrule
\textbf{$\boldsymbol{\omega_1}$} & & & & & \\
MAMO & 14332 & 25.7 & 23.5 & 22.4 & 28.4 \\
No UABS & 4236 & 100.0 & - & - & - \\
\hline\hline
\textbf{$\boldsymbol{\omega_2}$} & & & & & \\
MAMO & 19286 & 50.3 & 12.3 & 19.9 & 17.5 \\
No UABS & 11418 & 100.0 & - & - & - \\
\hline\hline
\textbf{$\boldsymbol{\omega_3}$} & & & & & \\
MAMO & 8974 & 49.2 & 14.1 & 16.6 & 20.1 \\
No UABS &  6316 & 100.0 & - & - & - \\
\bottomrule
\end{tabular}%
}
\label{tab:load_distribution}
\end{table}

%% file: Chapters/8_Conclusions.tex
\section{Conclusions}
\label{sec:conclusions}

This work addressed the challenge of adaptive trajectory design for \ac{UABS} fleets by developing an online \ac{MADRL} framework that can generalize across diverse operational environments. The proposed \ac{MAMO} framework introduces a meta-advisor that captures shared structural knowledge from multiple service areas and takeoff configurations. By combining this meta-advisor with a task-specific model, we enable agents to navigate the exploration-exploitation tradeoff effectively without the need for retraining when facing new scenarios. Moreover, a novel advisor rejection strategy allows agents to protect model training from potentially task-mismatched advisor suggestions. This strategy ensures both rapid multi-task adaptation and robustness to task heterogeneity.

Validation through simulations of realistic Bologna city districts confirms that \ac{MAMO} provides superior adaptability compared to traditional methods. 
Specifically, our results show that \ac{MAMO} reduces the first successful episode metric by up to 60\% compared to $\epsilon$-greedy baselines, while maintaining consistent performance across diverse tasks. The ablation study validates the necessity of the override mechanism, particularly in heterogeneous scenarios where the meta-advisor alone struggles to generalize.
From a network performance perspective, the trained \ac{UABS} fleet, cooperating with the existing \ac{MBS},  provides substantial improvements in quality of service compared to the case where no drones are deployed.
By facilitating coordinated and safe online learning, this approach offers a scalable solution for integrating autonomous aerial nodes into 6G vehicular networks.

%% file: Chapters/4_RRM_system.tex
\label{sec:RRM_model}
We consider radio \acp{RU}
spanning three dimensions: time, frequency, and space.
In the frequency domain, the bandwidth is subdivided into subcarriers with a spacing of $\Delta f$. A \ac{RU} consists of $N_{\rm sub}$ consecutive subcarriers, resulting in a total bandwidth of $B_{\rm RU} = \Delta f \cdot N_{\rm sub}$.
In the time domain, transmission is divided into slots of duration $T_{\rm slot}$. This duration depends on $\Delta f$, such that a single \ac{RU} accommodates 14 \ac{OFDM} symbols.

Assuming the \ac{RRM} algorithm executes with a periodicity of $\Delta t$, and given a total system bandwidth $B_{\rm sys}$, the total number of \acp{RU} available for scheduling within one \ac{RRM} period is:
\begin{equation}
W = \frac{B_{\rm sys}}{B_{\rm RU}}\cdot\frac{\Delta t}{ T_{\rm slot}}.
\end{equation}
Furthermore, the spatial domain may provide orthogonality among resources, increasing capacity and mitigating interference. This spatial separation is implemented via beamforming at both the \acp{MBS} and \ac{UABS}, establishing directional links to the \acp{GUE}.
Specifically, the \ac{UABS} generates a fixed footprint on the ground composed of $N_{\rm beam} = 9$ circular beams arranged in a $3\times3$ squared non-overlapping grid. As these beams are active simultaneously without overlap, full frequency reuse is possible across beams.
Conversely, \ac{MBS}-\ac{GUE} links are susceptible to building clutter and \ac{NLoS} conditions; thus, perfect spatial separation is not guaranteed. Consequently, while \ac{MBS} directional beams provide improved receiving gain, they do not support the full reuse architecture employed by the \ac{UABS}.

\label{sec:rrm_algo}
\input{Algo/UL_algo}
The primary objective of the \ac{RRM} algorithm is to maximize the number of served users, weighted by their priority to enforce service continuity. This is achieved by jointly optimizing user association (for both \ac{MBS} and \ac{UABS} fleets) and resource scheduling. 
To make this paper self-contained, the algorithm procedure is here briefly discussed; a more complete explanation can be found in~\cite{10457546}.
The problem is formulated as a centralized, multi-\ac{UABS} and multi-\ac{MBS} \ac{ILP}. 
At each time step $t$, the optimization runs on the following input variables:
(i) $k_{g,j_u}$ indicates whether vehicle $g$ is covered by beam $j_u \in \mathcal{K}_u$ of \ac{UABS} $u$, with $\mathcal{K}_u$ the set of beams with cardinality $N_{\rm beam}$; (ii) $I_{g,m,u}$ and $I_{g,u,m}$ are binary variables capturing potential interference topology, for instance, $I_{g,m,u}=1$ implies $g$ is in range of both $m$ and $u$, so a transmission from \ac{GUE} $g$ directed toward \ac{MBS} $m$ may interfere with \ac{UABS} $u$. A similar implication holds when $I_{g,u,m}=1$; (iii) $r_{g,u}, r_{g,m}, r_{u,m}$ and $r_{g,m,u}^I, r_{g,u,m}^I$ represents the per-\ac{RU} rate achieved between transmitters and receivers in, respectively, absence or in presence of interference.

The \ac{ILP} solution determines the network configuration through several decision variables:
(i) $\lambda_{g,m}, \lambda_{g,u} \in \{0,1\}$ represents user association, i.e. if \ac{GUE} $g$ is associated to \ac{MBS} $m$ or \ac{UABS} $u$. Then, $\lambda_{u,m}$ controls the activation of the backhaul between \ac{UABS} $u$ and \ac{MBS} $m$; 
(ii) $w_{g,m}, w_{g,u} \in \{0,W\}$ are integer variables representing the number of \acp{RU} allocated to \acp{GUE} $g$ for uplink transmission, while $w_{u,m} \in \{0,W\}$ the resources for backhaul; 
(iii) $e_{j_u}$, with $j_u \in \mathcal{K}_u$ are binary variables tracking the \ac{UABS} beam activation.
(iv) $\iota_{g,m,u}, \iota_{g,u,m}, \iota_{m,u}, \iota_{u,m} \in \{0,1\}$ indicate the true interference relationship given a potential user association decision. For example, $\iota_{g,m,u} = 1$ if a transmission from $g$ to $m$ actually suffers interference from a simultaneous service at \ac{UABS} $u$. As a result, the resource budget is computed assuming interference-limited rates. 
(v) $\psi_{g}$ is the binary variable representing the successful upload of the packet sent by \ac{GUE} $g$. Moreover, auxiliary variables $\psi_{g,u}=\psi_{g}\cdot\lambda_{g,u}$ and $\psi_{g,m}=\psi_{g}\cdot\lambda_{g,m}$ track the serving \ac{BS}.

These variables are governed by the constraints \eqref{u:demand}-\eqref{u:interf-m-on-gu}. Constraints \eqref{u:demand} and \eqref{u:demand-sinr-um} ensure demand $D_g$ is met by aggregating the rates of assigned \acp{RU}, accounting for both interference-free and interference-limited regimes, respectively. Resource limits are enforced by \eqref{u:res_limit_mbs} and \eqref{u:res_limit_uabs}, which cap the \acp{RU} assigned by \acp{MBS} and \acp{UABS}, strictly accounting for backhaul overhead. 
Constraint \eqref{u:capacity_backhaul} enforces uplink transmission limits while guaranteeing sufficient backhaul capacity to support the aggregate vehicular traffic transmitted by each \ac{UABS}.
Constraints \eqref{u:beam_uabs_lim} and \eqref{u:beam_uabs_lim2} restrict the active beams per \ac{UABS} to $N_{\text{beam}}$, while constraints \eqref{u:one_base}--\eqref{u:one_base_uabs} ensure a vehicle connects to exactly one \ac{BS} at a time.
The logic linking potential interference to realized interference is enforced by \eqref{u:interf-u-on-m}--\eqref{u:interf-m-on-gu}. Specifically, 
the constraints~\eqref{u:interf-u-on-m} and~\eqref{u:interf-m-on-u} verify if there is at least one effective interferer on the \ac{MBS} that is connected to a \acp{UABS} $u$ or vice versa, respectively. Then,
\eqref{u:interf-u-on-gm} determines if a \ac{UABS} $u$ interferes with an active link between $g \text{ and } m$, while \eqref{u:interf-m-on-gu} checks if an \ac{MBS} $m$ interferes with an active link between $g \text{ and } u$.

%% file: Algo/UL_algo.tex
\begin{algorithm*}[tbh]
    \centering
    \footnotesize
    
    \begin{subequations}
    \begin{equation}
        \mathcal{P}: \max \sum_{g \in \mathcal{G}} \left(\psi_{g} \cdot p_{g} \right)
        \label{u:obj}
    \end{equation}
    \hfill
    
    \begin{equation}
    \begin{split}
        \textbf{s.t.: } \sum_{m \in \mathcal{M}} w_{g, m} r_{g, m} \Delta t +\sum_{u \in \mathcal{U}} \sum_{j_u \in \mathcal{K}_u} k_{g, j_u}  w_{g, u} r_{g, u} \Delta t \geq \psi_{g} D_g   \text{, } \forall g \in \mathcal{G} \label{u:demand} 
    \end{split}
    \end{equation}
    \centering
    \begin{equation}
    \begin{split}
    \sum_{m \in \mathcal{M}} w_{g, m} r^{\rm I}_{g, m, u} \Delta t +\sum_{u \in \mathcal{U}} \sum_{j_u \in \mathcal{K}_u} k_{g, j_u}  w_{g, u} r^{\rm I}_{g, u, m} \Delta t \geq \Big(\!\sum_{m \in \mathcal{M}}\!\iota_{g,m,u}\!+\!\sum_{u \in \mathcal{U}}\!\iota_{g,u,m}\Big)D_g \text{, }\\ \forall g \in \mathcal{G},  \forall u \in \mathcal{U}, \forall m\!\in\!\mathcal{M} \label{u:demand-sinr-um}
    \end{split}
    \end{equation}
    
    \begin{minipage}{0.47\textwidth}
        \begin{equation}
            \sum_{g \in \mathcal{G}} w_{g, m} +  \sum_{u \in \mathcal{U}} w_{u, m} \leq W\text{, }\forall m \in \mathcal{M} \label{u:res_limit_mbs}
        \end{equation}
    \end{minipage}
    \hfill
    \begin{minipage}{0.47\textwidth}
        \begin{equation}
            \sum_{g \in \mathcal{G}} k_{g, j_u} w_{g, u} + \sum_{m \in \mathcal{M}} w_{u,m} \leq W   \text{, }\forall u \in \mathcal{U} \text{, } \forall j_u \in \mathcal{K}_u \label{u:res_limit_uabs}
        \end{equation}
    \end{minipage}
    
    \begin{minipage}{0.47\textwidth}
        \begin{equation}
            \sum_{g \in \mathcal{G}} \sum_{j_u \in \mathcal{K}_u} w_{g, u} k_{g, j_u} r_{g, u} \leq \sum_{m \in \mathcal{M}} r_{u, m} w_{u, m} \text{, }\forall u \in \mathcal{U} \label{u:capacity_backhaul}
        \end{equation}
    \end{minipage}
    \hfill
    \begin{minipage}{0.47\textwidth}
        \begin{equation}
            \sum_{j_u \in \mathcal{K}_u} e_{j_u} \leq N_{\rm beam} \text{, } \forall u \in \mathcal{U} \label{u:beam_uabs_lim}
        \end{equation}
    \end{minipage}

    \begin{minipage}{0.47\textwidth}
        \begin{equation}
            \sum_{g \in \mathcal{G}} w_{g, u} k_{g,j_u} \leq e_{j_u} W \text{, } \forall u \in \mathcal{U} \text{, }\forall j_u \in \mathcal{K}_u \label{u:beam_uabs_lim2}
        \end{equation}
    \end{minipage}
    \hfill
    \begin{minipage}{0.47\textwidth}
        \begin{equation}
            \sum_{m \in \mathcal{M}} \lambda_{g, m} + \sum_{u \in \mathcal{U}} \lambda_{g,u} \leq 1  \text{, }\forall g \in \mathcal{G} \label{u:one_base}
        \end{equation}
    \end{minipage}

    \begin{minipage}{0.47\textwidth}
        \begin{equation}
             w_{g, m} \leq \lambda_{g, m} W \text{, }\forall m \in \mathcal{M}, \forall g \in \mathcal{G} \label{u:one_base_mbs}
        \end{equation}
    \end{minipage}
    \hfill
    \begin{minipage}{0.47\textwidth}
        \begin{equation}
           w_{g, u} \leq \lambda_{g, u} W \text{, }\forall u \in \mathcal{U} \text{, } \forall g \in \mathcal{G} \label{u:one_base_uabs}
        \end{equation}
    \end{minipage}

    \begin{minipage}{0.47\textwidth}
        \begin{equation}
            \iota_{m,u} \geq \sum_{g \in \mathcal{G}}\frac{I_{g,u,m} \lambda_{g,u}}{I_{g,u,m}} \text{, }\forall u \in \mathcal{U} , \forall m \in \mathcal{M}\label{u:interf-u-on-m}            
        \end{equation}
    \end{minipage}
    \hfill
    \begin{minipage}{0.47\textwidth}
        \begin{equation}
            \iota_{u,m} \geq \sum_{g \in \mathcal{G}}\frac{I_{g,m,u} \lambda_{g,m}}{I_{g,m,u}} \text{, }\forall m \in \mathcal{M}, \forall u \in \mathcal{U} \label{u:interf-m-on-u}
        \end{equation}
    \end{minipage}

    \begin{minipage}{0.47\textwidth}
        \begin{equation}
            \iota_{g,m,u} \geq \lambda_{g,m} + \iota_{m,u} - 1\text{, } \forall g \in \mathcal{G} \text{, }\forall u \in \mathcal{U}, \forall m \in \mathcal{M} \label{u:interf-u-on-gm}
        \end{equation}
    \end{minipage}
    \hfill
    \begin{minipage}{0.47\textwidth}
        \begin{equation}
            \iota_{g,u,m} \geq \lambda_{g,u} + \iota_{u,m} - 1 \text{, }\forall g \in \mathcal{G} \text{, }\forall m \in \mathcal{M}, \forall u \in \mathcal{U} \label{u:interf-m-on-gu}
        \end{equation}
    \end{minipage}
    \end{subequations}
    \vspace{1em}
    \caption{ILP formulation for the generalized RRM problem.}
    \label{alg:RRM_algorithm}
\end{algorithm*}
\normalsize